\documentclass[draftcls, 12pt,onecolumn,oneside]{IEEEtran}
\usepackage{pifont}
\usepackage{times,amsmath,cite,color,amssymb,graphicx,epsfig,cite,geometry,psfrag,subfigure,algorithm,color}
 \makeatletter
\newcommand{\Rmnum}[1]{\expandafter\@slowromancap\romannumeral #1@}
\makeatother

 \allowdisplaybreaks[4]

\begin{document}

\title{ Frequency Synchronization for Uplink Massive MIMO  Systems  }
\author{Weile Zhang, Feifei Gao, Shi Jin, and Hai Lin
\thanks{
W. Zhang is with the MOE Key Lab for Intelligent Networks and Network Security,
Xi'an Jiaotong University, Xi'an, Shaanxi, 710049, China. (email:
wlzhang@mail.xjtu.edu.cn)
 }
 \thanks{F. Gao is with the State Key Laboratory of
Intelligent Technology and Systems, Tsinghua National Laboratory for
Information Science and Technology, Department of Automation,
Tsinghua University, Beijing, 100084, China (e-mail:
feifeigao@ieee.org).}
\thanks{S. Jin is with the National Communications Research Laboratory, Southeast University, Nanjing 210096, P. R. China (email: jinshi@seu.edu.cn).}
\thanks{H. Lin is with the Department of Electrical and Information Systems, Osaka
Prefecture University, Osaka, Japan. (email: hai.lin@ieee.org)}
}

 \maketitle

\vspace{-15mm}

 \begin{abstract}
In this paper, we propose a frequency synchronization scheme for multiuser orthogonal frequency division multiplexing (OFDM) uplink with a large-scale uniform linear array (ULA) at base station (BS) by exploiting the angle information of users.
 Considering that the incident signal at BS from each user can be restricted within a certain angular spread, the proposed scheme could perform carrier frequency offset (CFO) estimation for each user individually through a \textit{joint spatial-frequency alignment} procedure and can be completed efficiently with the aided of  fast Fourier transform (FFT).
A multi-branch receive beamforming is further designed to yield an equivalent single user transmission model for which the conventional single-user channel estimation and data detection can be carried out.
To make the study complete, the theoretical performance analysis of the CFO estimation is also conducted.
We further develop a user grouping  scheme to deal with the unexpected scenarios that some users may not be separated well from the spatial domain.
Finally, various numerical results are provided to verify the proposed studies.
 \end{abstract}


\section{Introduction}

Large-scale multiple-input multiple-output (MIMO) or ``massive MIMO'' systems have drawn considerable interests from both academia and industry~\cite{Marzetta10,Choi14,RusekSPM,LarssonCM}. In a typical massive MIMO system, the base station (BS) is equipped with hundreds of antennas to simultaneously communicate with only a few number of single-antenna users at the same frequency band.
Theoretically, such a massive MIMO system can almost perfectly relieve the inter-user interference in multiuser MIMO systems with simple linear transceivers. Other advantages, such as high spectral efficiency or robustness, also make massive MIMO a promising technique for the next generation  wireless systems.

However, all these potential gains of massive MIMO systems rely heavily on perfect frequency synchronization among multiple users, which is rather challenging due to the coexistence of multiple CFOs at the BS.
Both the multiuser CFO estimation and compensation problem are quite different from those in conventional single-user communication systems~\cite{Schmidl97, Minn03, Yao05, Lmai:TVT14, ZhangTSP14,ZhangTWC16}.
In the past decade, there have been a number of works  on the frequency synchronization for the conventional small-scale spatially multiplexed multiuser MIMO systems. For example, Besson and Stoica made the first attempt
on multiple CFOs estimation~\cite{Besson} for flat fading channels.
 A joint CFO and channel estimation for multiuser cyclic-prefix (CP)
MIMO orthogonal frequency division multiplexing (OFDM) systems was developed in~\cite{JChen08} by exploiting the
maximum likelihood (ML) criterion.
Another suboptimal estimation algorithm was proposed in~\cite{YWuEURAPSIP} based on long constant amplitude zero autocorrelation (CAZAC) training sequence.
The optimal set of training Zadoff-Chu sequences for multiple CFOs estimation was studied in~\cite{Tsai13}.
 The work in~\cite{DuttaTVT15} discretized the continuous-valued CFO parameters into a discrete set of bins and invoked the detection theory for CFO estimation.

The above works~\cite{Besson,JChen08,YWuEURAPSIP,Tsai13,DuttaTVT15} for small-scale MIMO usually considered very few spatially multiplexed users.
With the increase of the number of users, for example, in massive MIMO system, these methods may suffer from substantial performance degradation and may not be applied.
Recently, the authors in~\cite{Cheng13} proposed an approximation to the joint maximum likelihood (ML) estimation for massive MIMO systems. However, the estimator~\cite{Cheng13} is designed for frequency flat channels and requires a multi-dimensional grid search.
The work in~\cite{MukherjeeTVT} studied the impact of imperfect CFO estimation/compensation on the performance of  zero-forcing (ZF) receiver for massive MIMO uplink transmissions.
 The authors in~\cite{Mukherjee15} addressed the CFO estimation for time division duplexed (TDD) massive MIMO system and proposed a simple uplink training scheme. The time-orthogonal training sequences from different users are guarded by zero interval in~\cite{Mukherjee15}, such that the optimality criterion can be satisfied.
However, the CFO estimates at BS must be fed back to multiple users over an error free control channel in~\cite{Mukherjee15}, and thus the users can correct their oscillators accordingly before the data transmission.
\textcolor{black}{
 The authors further proposed a spatially averaged periodogram based method for CFO estimation with a constant envelope pilot signal in~\cite{MukherjeeGLO16}.
The impact of frequency selectivity on the performance of CFO impaired single-carrier massive MU-MIMO uplink has been further studied in~\cite{Mukherjee16}}.
More recently, the authors in~\cite{Zhang15sub} proposed a blind frequency synchronization scheme for multiuser OFDM uplink with large number of antennas at the receiver. The scheme in~\cite{Zhang15sub} is blind and thus has the potential benefit of saving system resources. However, it may need a large number of blocks to obtain a satisfactory estimation performance, especially in the low signal-to-noise (SNR) regime.

 An interesting array signal processing aided massive MIMO transmission scheme was proposed in~\cite{Gaoarxiv,GaoAccess},
where the angle information of the user is exploited to separate users and simplify the subsequent signal processing. It is shown that users can be grouped according to their angle information to release the inter-user interference, which can be named as angle division multiple access (ADMA). However, the work mainly focused on the channel estimation but does not provide the solution for frequency synchronization.

Motivated by~\cite{Gaoarxiv,GaoAccess}, in this paper, we design a new frequency synchronization scheme for multiuser OFDM uplink with a massive uniform linear array (ULA) at BS by exploiting the angle information of users. We  consider that the incident signals at BS from each user can be restricted within a certain angular spread, since the massive BS is usually elevated at a very high altitude with few surrounding scatterers~\cite{Adhikary13,You15,Sun15,Gaoarxiv,GaoAccess}.
With sufficient spatial dimensions, the multiuser interference (MUI) effect can be substantially mitigated in angle domain via beamforming with the steering vectors pointing to the direction of the incoming signals. The proposed scheme could perform CFO estimation for each user individually through a \textit{joint spatial-frequency alignment} procedure. A multi-branch beamforming is designed for each user and yield an equivalent single user transmission model for which the conventional single-user channel estimation and data detection can be further carried out.
Moreover, the proposed scheme can be applied efficiently with the aided of fast Fourier transform (FFT) operation and therefore is of low complexity. To make the study complete, we also conduct the theoretical performance analysis for CFO estimation.  Considering some users may not be separated well from the spatial domain, we further develop a user grouping based scheme, in which the users with similar DOAs are grouped together and a joint CFO estimation and data detection method is designed for each group.  Finally, we  provide numerical results to verify the proposed studies.

The rest of this paper is organized as follows. The system model is
described in Section II. The  proposed frequency synchronization scheme is presented in Section III. The improved scheme with user grouping is further developed in Section IV.
 Simulation results are
provided in Section V and conclusions are drawn in Section VI.

\textit{Notations:} Superscripts $(\cdot)^*$, $(\cdot)^T$,
$(\cdot)^H$, $(\cdot)^\dag$ and $E[\cdot]$ represent conjugate, transpose, Hermitian, pseudo-inverse
and expectation, respectively; ${\bf j}=\sqrt{-1}$ is the imaginary
unit;  $\|\cdot\|$ denotes the Frobenius norm operator; for a vector
$\boldsymbol x$, diag$(\boldsymbol x)$ is a diagonal matrix with
main diagonal of $\boldsymbol x$;
$\otimes$ stands for the Kronecker product; $\textrm{vec}(\cdot)$ is the vectorization operator;
 ${\mathbb C}^{m\times n}$ defines the vector space of all
$m\times n$ complex matrices; ${\bf I}_N$ is the $N\times N$
identity matrix; ${\bf 0}$ represents an all-zero matrix with
appropriate dimension;  $\textrm{Tr}(\cdot)$ denotes the trace
operation. Parts of this paper have been presented in a conference paper~\cite{ZhangGLO16}.

\section{System Model}

\begin{figure}[t]
\begin{center}
\includegraphics[width=80mm]{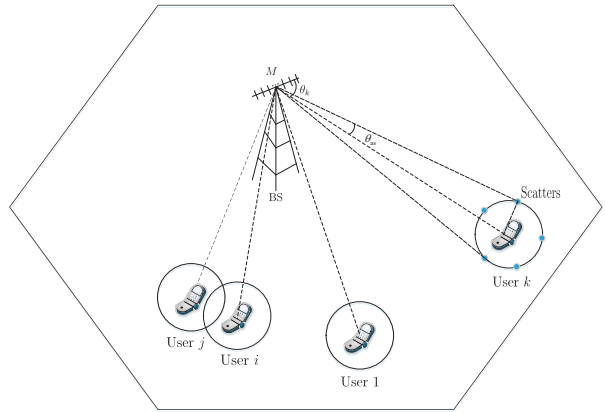}
\end{center}
\vspace{-5mm}
\caption{ System of massive multiuser uplink transmissions. Users are randomly distributed and surrounded by $P$ local scatterers. The mean DOA and angular spread of the $k\rm{th}$ user are $\theta_k$ and $\theta_{\rm as}$, respectively.}
\end{figure}

Consider a multiuser OFDM uplink system that consists of $K$ distributed single-antenna users and one BS with $M\gg 1$ antennas in the form of ULA, as shown in Fig. 1.  The total number of subcarriers is $N$ and we assume prefect time synchronization among all users for the timing being. Denote the normalized CFO between the $k\rm{th}$ user and the BS by $\phi_k$,
\textcolor{black}{which is the ratio between the real CFO and the subcarrier spacing.
Let
$\Delta f_k$ and $T_s$ stand for real CFO of the $k\rm{th}$ user and the sampling  interval of symbol-rate, respectively.
We have $2\pi \phi_k = 2\pi NT_s \Delta f_k $~\cite{Morelli}, which represents the CFO-induced phase shift over one OFDM block. }
We consider the fractional CFO,  i.e., $|\phi_k|<0.5$, which should be sufficient in the multiuser uplink~\cite{Morelli,Sun09}.

The classical one-ring channel propagation model is adopted \cite{Adhikary13,You15,Sun15}, where each user is surrounded by a ring of $P\gg 1$ local scatterers. The multi-tap channel matrix between the $k\rm{th}$ user and the BS can be  modeled as
an $L\times M$ matrix
${\bf H}_k = \big[ {\bf h}^{(k)}_1,{\bf h}^{(k)}_2,\cdots, {\bf h}^{(k)}_L    \big]^T$,
where the $l$-$\rm{th}$ tap channel vector ${\bf h}^{(k)}_l\in\mathbb{C}^{M\times 1}$ is composed of $P$ rays and can be expressed as:
\begin{align}\label{equ25}
{\bf h}^{(k)}_l =  \sum_{p=1}^P \alpha_{l,p,k} {\boldsymbol a}(\theta_{l,p,k} ) .
\end{align}
Here, $\theta_{l,p,k}$ represents the DOA of the $p\rm{th}$ ray in the $l$-$\rm{th}$ channel tap, \textcolor{black}{while $\alpha_{l,p,k}\sim \mathcal{CN}(0, \sigma_{h,l}^2/P)$ }represents the corresponding complex gain and is independent from each other~\cite{You15}.
\textcolor{black}{Besides, $\sigma_{h,l}^2$, $l=1,2,\cdots,L$, models the power delay profile (PDP) of the propagation channel.  We consider $\sum_{l=1}^L\sigma_{h,l}^2=1 $ such that the total channel gain of each user at one receive antenna is normalized. }
 Moreover, ${\boldsymbol a}(\theta_{l,p,k} ) \in\mathbb{C}^{M\times 1}$ is the steering vector and has the form
\begin{align}
{\boldsymbol a}(\theta_{l,p,k} )  = \big [1, {\rm e}^{-{\bf j}\chi\cos\theta_{l,p,k}},\cdots, {\rm e}^{-{\bf j}\chi(M-1)\cos\theta_{l,p,k}} \big]^T,
\end{align}
where $\chi = \frac{2\pi d}{\lambda}$, $d$ is the antenna spacing, $\lambda$ denotes the signal wavelength.
Similar to \cite{Adhikary13,You15,Sun15,Gaoarxiv,GaoAccess}, we consider that the incident rays of each user can be constrained within a certain angular spread $\theta_{\rm as}$; Namely,  the incident angles of the $k\rm{th}$ user is limited in the DOA region $(\theta_k-\theta_{\rm as}, \theta_k + \theta_{\rm as})$ with uniform distribution, where  $\theta_k$ represents the mean DOA of the $k\rm{th}$ user.
\textcolor{black}{Generally, this assumption has been supported in the following typical scenario~\cite{GaoAccess}:  BS equipped with a large number of antennas is always elevated at a very high altitude, say on the top of a high building, a dedicated tower, or an unmanned aerial vehicle platform, such that there are few surrounding scatterers. Hence, the angular spread seen by BS is quite small.}

Define ${\bf F}$ as the $N\times N$ normalized DFT matrix with its ($i,j$)-$\rm{th}$ entry ${\bf F}_{i,j}=\frac{1}{\sqrt{N}}{\rm e}^{-{\bf j}\frac{2\pi (i-1)(j-1)}{N}}$. Let ${\bf F}_L$ stand for the submatrix of ${\bf F}$ containing the first $L$ column vectors and ${\bf E}(\phi_k)$ represent the $N\times N$ diagonal phase rotation matrix:
${\bf E}(\phi_k) = \textrm{diag}(1, {\rm e}^{{\bf j}\frac{2\pi\phi_k}{N}},\cdots, {\rm e}^{{\bf j}\frac{2\pi(N-1)\phi_k}{N}}  )$.

\begin{figure}[t]
\begin{center}
\includegraphics[width=110mm]{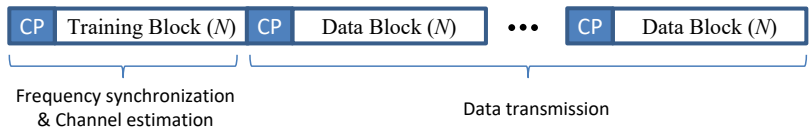}
\end{center}
\vspace{-8mm}
\caption{\textcolor{black}{The uplink communication strategy. All $K$ users simultaneously transmit the training and data blocks to BS. }
}
\end{figure}

Each user transmits a whole training block at the beginning of the uplink frame, which is then followed by a number of data blocks.
\textcolor{black}{The uplink communication strategy is described in Fig. 2.}
Denote ${\bf x}_k = \big[x_k(0), x_k(1), \cdots, x_k(N-1)\big]^T$ as the frequency-domain training block transmitted from the $k\rm{th}$ user.
 Then, the received time-domain training signal after CP removal can be written as the following $N\times M$ matrix:
\begin{align}\label{bfY}
{\bf Y} = \sum_{k=1}^K  {\bf E}(\phi_k) {\bf B}_k {\bf H}_k + {\bf N},
\end{align}
where
${\bf B}_k = \sqrt{N} {\bf F}^H \textrm{diag}({\bf x}_k){\bf F}_L \in\mathbb{C}^{N\times L}$,
 and ${\bf N}$ denotes the corresponding $N\times M$ additive white Gaussian noise (AWGN) matrix. We assume each element of ${\bf N}$ follows i.i.d. complex Gaussian distribution with variance $\sigma_n^2$, i.e., $E[{\bf N}{\bf N}^H]=M\sigma_n^2 {\bf I}_N$.

Denote ${\bf s}_i^{(k)} \in \mathbb{C}^{N\times 1}$ as the frequency-domain data symbol vector transmitted from the $k\rm{th}$ user in the $i$-$\rm{th}$ data block.
The corresponding received signal at BS can be expressed as the following $N\times M$ matrix:
\begin{align}
{\bf Y}_{i}^{\rm d} =  \sqrt{N} \sum_{k=1}^K \eta_i(\phi_k) {\bf E}(\phi_k) {\bf F}^H \textrm{diag}({\bf s}_i^{(k)}){\bf F}_L {\bf H}_k + {\bf N}_i^{\rm d},
\end{align}
where $\eta_i(\phi_k) = {\rm e}^{{\bf j}\frac{2\pi i (N+N_{\rm cp})\phi_k}{N}}$ represents the accumulative phase rotation introduced by the CFO of the $k\rm{th}$ user with $N_{\rm cp}$ being the length of CP, and ${\bf N}_i^d \in \mathbb{C}^{N\times M}$ denotes the corresponding noise matrix with $E[{\bf N}_i^{\rm d}({\bf N}_i^{\rm d})^H]=M\sigma_n^2 {\bf I}_N$.

\section{Joint Frequency Synchronization and Beamforming}

\subsection{CFO Estimation}

 Denote ${\bf P}_{{\bf B}_k} = {\bf B}_k ({\bf B}_k^H {\bf B}_k)^{-1}{\bf B}_k^H$ as the orthogonal projection operator onto the space spanned by the columns of ${\bf B}_k$, and define   ${\bf P}_{{\bf B}_k}^\bot = {\bf I}_N - {\bf P}_{{\bf B}_k}$. In order to guarantee the identifiability of the CFO estimation for the $k\rm{th}$ user,  we consider the training sequence is designed such that ${\bf P}_{{\bf B}_k}^\bot {\bf E}(\tilde\phi) {\bf B}_k \ne {\bf 0}$ holds for any $\tilde\phi\ne 0$. Meanwhile, assume ${\bf P}_{{\bf B}_k}^\bot {\bf E}(\tilde\phi) {\bf B}_q \ne {\bf 0}$ holds for any $\tilde\phi$ and $k\ne q$.

The proposed multiuser frequency synchronization scheme is motivated by the following intuitive observation. With sufficient number of receive antennas, i.e., the sufficient spatial dimension,
the steering vectors pointing to distinct DOAs would become nearly orthogonal to each other, which implies that we can simply employ the steering vector pointing to the DOA of the $k\rm{th}$ user as the beamforming vector to substantially suppress the interference from the other users in angle domain~\cite{Gaoarxiv,GaoAccess}. This will yield an approximate single-user received signal model for the $k\rm{th}$ user, where many conventional single-user frequency synchronization algorithm can be carried out.

Let us first assume that the angular spread of all simultaneous users are disjoint through some proper user schedule strategy and
consider $M\to \infty$ for the first glance. It can be readily checked that
the steering vectors would become orthogonal to each other for any two distinct DOAs; namely ${\boldsymbol a}(\theta_1)^H {\boldsymbol a}(\theta_2)/M = 0$ holds for any $\theta_1\ne \theta_2$~\cite{You15,Gaoarxiv}.
Hence, for the $k\rm{th}$ user, when the angular spread of users are disjoint,  the beamforming vector ${\boldsymbol a}(\tilde\theta)$ with \textit{spatial alignment}, i.e., $\tilde\theta\in (\theta_k-\theta_{\rm as}, \theta_k + \theta_{\rm as})$,  will eliminate the signals from the other users; namely, there are ${\bf H}_k {\boldsymbol a}^*(\tilde\theta)\ne {\bf 0}$ and ${\bf H}_q {\boldsymbol a}^*(\tilde\theta)= {\bf 0}$ for any $q\ne k$.  We then obtain
 \begin{align}\label{equ8}
 {\bf Y}{\boldsymbol a}^*(\tilde\theta) = {\bf E}(\phi_k) {\bf B}_k {\bf H}_k {\boldsymbol a}^*(\tilde\theta) + \underbrace{ \sum_{q\ne k} {\bf E}(\phi_q) {\bf B}_q {\bf H}_q  {\boldsymbol a}^*(\tilde\theta) }_{\textrm{MUI term} \hspace{1pt}\simeq \hspace{1pt} {\bf 0} } + {\bf N}{\boldsymbol a}^*(\tilde\theta),
 \end{align}
which gives an equivalent frequency asynchronous  single-user received signal model for the $k\rm{th}$ user.
\textcolor{black}{
The MUI term of (\ref{equ8}) approaching zero arises from the following observation:
The $l\rm{th}$ entry of ${\bf H}_q  {\boldsymbol a}^*(\tilde\theta)$, $l=1,2,\cdots,L$, equals $({\bf h}^{(q)}_l)^T {\boldsymbol a}^*(\tilde\theta) = \sum_{p=1}^P \alpha_{l,p,q} {\boldsymbol a}^T(\theta_{l,p,q}){\boldsymbol a}^*(\tilde\theta) $.
Once the angular spreads of the $k\rm{th}$ and the $q\rm{th}$ user are disjoint,  ${\boldsymbol a}(\tilde\theta)$ is nearly orthogonal to any ${\boldsymbol a}(\theta_{l,p,q})$ with sufficient large $M$~\cite{You15,Gaoarxiv}. }

We can further perform proper CFO compensation of (\ref{equ8}) for purpose of \textit{frequency alignment} as
 \begin{align}\label{equ9}
 {\bf E}(\phi_k)^H{\bf Y}{\boldsymbol a}^*(\tilde\theta) = {\bf B}_k {\bf H}_k {\boldsymbol a}^*(\tilde\theta) +  {\bf E}(\phi_k)^H{\bf N}{\boldsymbol a}^*(\tilde\theta).
 \end{align}
This implies that with the above spatial and frequency alignment, an equivalent frequency synchronous single-user received signal model will be obtained at BS corresponding to the $k\rm{th}$ user.
 The resultant signal in (\ref{equ9}) would exactly lie in the column space of ${\bf B}_k$ in the absence of noise.
It can be observed that, if $\tilde\theta\notin (\theta_k-\theta_{\rm as}, \theta_k + \theta_{\rm as})$ or if $\tilde\phi\ne \phi_k$, then the multiuser interference or frequency asynchronism would deviate the column space of   ${\bf E}(\tilde\phi)^H{\bf Y}{\boldsymbol a}^*(\tilde\theta)$  from that of $\textrm{Span}({\bf B}_k)$.

Based on the above observations, the frequency synchronization for the $k\rm{th}$ user can be designed as the following \textit{joint spatial-frequency alignment} procedure.
Specifically, the joint CFO and DOA estimation for the $k\rm{th}$ user can be obtained by minimizing the cost function:
\begin{align}\label{costfunc}
 C_k(\tilde\theta_k,\tilde\phi_k) = \frac{ \int_{\tilde\theta_k-\theta_{\rm as}}^{{\tilde\theta_k+ \theta_{\rm as}}} \|{\bf P}_{{\bf B}_k}^\bot{\bf E}^H(\tilde\phi_k){\bf Y}{\boldsymbol a}^*(\tilde\theta)\|^2  d\tilde\theta} {  \int_{\tilde\theta_k-\theta_{\rm as}}^{{\tilde\theta_k+ \theta_{\rm as}}} \| {\bf Y}{\boldsymbol a}^*(\tilde\theta) \|^2  d\tilde\theta  },
\end{align}
where the trial DOA $\tilde\theta_k$ and the trial CFO $\tilde\phi_k$ target at the spatial and frequency alignment, respectively.
Mathematically, the joint CFO and DOA estimation for the $k\rm{th}$ user can be expressed as
\begin{align}
& \big\{\hat\theta_k, \hat\phi_k\big\} =  \arg \min_{\tilde\theta_k,\tilde\phi_k}  C_k(\tilde\theta_k,\tilde\phi_k).   \label{equ1}
\end{align}

We further obtain the following important \textit{Property}:

\textit{Property:} At high SNR region, the cost function in (\ref{equ1}) achieves its minimum only in the case of both spatial and frequency alignment, i.e., $\tilde\theta_k=\theta_k$ and $\tilde\phi_k=\phi_k$.
\begin{proof}
When $\tilde\theta_k = \theta_k$, all DOAs from the $k\rm{th}$ user will be included in the integration of (\ref{costfunc}) and then the beamforming vectors ${\boldsymbol a}(\tilde\theta)$ in (\ref{costfunc}) will eliminate the signals from all users other than the $k\rm{th}$ user.
In addition, when $\tilde\phi_k = \phi_k$, there is ${\bf E}(\tilde\phi_k)^H{\bf Y}{\boldsymbol a}^*(\tilde\theta)= {\bf B}_k{\bf H}_k {\boldsymbol a}^*(\tilde\theta) + {\bf E}(\tilde\phi_k)^H{\bf N}{\boldsymbol a}^*(\tilde\theta)$. As a result,   the cost function is expressed as
\begin{align}
 C_k(\theta_k,\phi_k) \simeq \frac{ \int_{\theta_k-\theta_{\rm as}}^{{\theta_k+ \theta_{\rm as}}} \|{\bf P}_{{\bf B}_k}^\bot {\bf E}(\phi_k)^H {\bf N}{\boldsymbol a}^*(\tilde\theta)\|^2  d\tilde\theta} {  \int_{\theta_k-\theta_{\rm as}}^{{\theta_k+ \theta_{\rm as}}} \| {\bf B}_k {\bf H}_k{\boldsymbol a}^*(\tilde\theta) \|^2  d\tilde\theta  } \simeq \frac{ 2\theta_{\rm as} M (N-L) \sigma_n^2} {  \int_{\theta_k-\theta_{\rm as}}^{{\theta_k+ \theta_{\rm as}}} \| {\bf B}_k {\bf H}_k{\boldsymbol a}^*(\tilde\theta) \|^2  d\tilde\theta  },
\end{align}
which tends to zero as SNR increases. On the other hand, the spatial and frequency misalignment includes the following four cases:
\begin{enumerate}
\item
In the case of $\tilde\theta_k =\theta_k$ and $\tilde\phi_k \ne \phi_k$, the numerator of the cost function $C_k(\theta_k,\tilde\phi_k)$ approximately equals
$\int_{\theta_k-\theta_{\rm as}}^{{\theta_k+ \theta_{\rm as}}} \|{\bf P}_{{\bf B}_k}^\bot {\bf E}(\phi_k-\tilde\phi_k) {\bf B}_k {\bf H}_k{\boldsymbol a}^*(\tilde\theta)\|^2  d\tilde\theta + 2\theta_{\rm as}M(N-L)\sigma_n^2$, which indicates that additional signal power will be included in the numerator of (\ref{costfunc}). Then there is $C_k(\theta_k,\tilde\phi_k) > C_k(\theta_k,\phi_k)$.
\item
When $0<|\tilde\theta_k - \theta_k|<2\theta_{\rm as}$,
the integration interval $(\tilde\theta_k\!-\!\theta_{\rm as}, \tilde\theta_k\!+\!\theta_{\rm as})$ intersects but not exactly coincides with the DOA region of the $k\rm{th}$ user. In this case,  not all of the signal components of the $k\rm{th}$ user are  included in the denominator of the cost function. Thus,  there is $C_k(\tilde\theta_k,\tilde\phi_k) > C_k(\theta_k,\phi_k)$.
\item
When $|\tilde\theta_k - \theta_q|>2\theta_{\rm as}$, $\forall q=1,2,\cdots,K$,  the integration interval $(\tilde\theta_k\!-\!\theta_{\rm as}, \tilde\theta_k\!+\!\theta_{\rm as})$ dose not intersect with the DOA region of any user. In this case,  the denominator of the cost function approximately equals $ \int_{\tilde\theta_k-\theta_{\rm as}}^{{\tilde\theta_k+ \theta_{\rm as}}} \| {\bf N}{\boldsymbol a}^*(\tilde\theta) \|^2 d\tilde\theta\simeq 2\theta_{\rm as} M N \sigma_n^2$ and the cost function becomes $(N-L)/N$. We could obtain $C_k(\tilde\theta_k,\tilde\phi_k) > C_k(\theta_k,\phi_k)$ at high SNR condition.
\item
When the integration interval intersects with the DOA region of one user (say the $q\rm{th}$ user) other than the $k\rm{th}$ user,  the numerator of the cost function $C_k(\tilde\theta_k,\tilde\phi_k)$ approximately equals
$\int_{\tilde\theta_k-\theta_{\rm as}}^{{\tilde\theta_k+ \theta_{\rm as}}} \|{\bf P}_{{\bf B}_k}^\bot {\bf E}(\phi_q-\tilde\phi_k) {\bf B}_q {\bf H}_q{\boldsymbol a}^*(\tilde\theta)\|^2  d\tilde\theta + 2\theta_{\rm as}M(N-L)\sigma_n^2$. In this case, the cost function dose not tend to zero as SNR increases.  Hence, $C_k(\tilde\theta_k,\tilde\phi_k) > C_k(\theta_k,\phi_k)$ holds at high SNR condition.
\end{enumerate}
From the above discussions, we know that at high SNR region, the cost function achieves its minimum only in the case of both spatial and frequency alignment. This completes the proof.
\end{proof}

The above \textit{Property} implies that, by finding the minimum point of (\ref{equ1}), we can obtain a valid estimation of both CFO and DOA for the $k\rm{th}$ user.
Note that directly solving the minimization problem (\ref{equ1}) requires two dimensional search with respect to both the trial CFO and DOA. \textcolor{black}{A low complexity iterative solution is further proposed by using the Taylor approximation.}
Specifically, denote the CFO estimation for the $k\rm{th}$ user in the $n\rm{th}$ iteration by $\hat\phi_k^{(n)}$. When $\tilde\phi$ is in the neighborhood of $\hat\phi_k^{(n)}$, we have the Taylor approximation of
${\bf E}(\tilde\phi) \simeq {\bf E}(\hat\phi^{(n)}_k)({\bf I}_N + {\bf D} \Delta\phi )$, where $\Delta\phi = \tilde\phi - \hat\phi^{(n)}_k$ and
 ${\bf D} = \frac{{\bf j}2\pi}{N} \textrm{diag}(0,1,\cdots, N-1)$.
We further denote ${\bf Y}^{(n)}_k =  {\bf E}^H(\hat\phi^{(n)}_k) {\bf Y}$, which corresponds to the signal matrix after the CFO compensation by $\hat\phi_k^{(n)}$.
Then, there is
\begin{align}
& \int_{\tilde\theta_k-\theta_{\rm as}}^{{\tilde\theta_k+ \theta_{\rm as}}} \|{\bf P}_{{\bf B}_k}^\bot {\bf E}(\tilde\phi) {\bf Y}{\boldsymbol a}^*(\tilde\theta)\|^2  d\tilde\theta \nonumber \\ =&
\int_{\tilde\theta_k-\theta_{\rm as}}^{{\tilde\theta_k + \theta_{\rm as}}} \| {\bf P}_{{\bf B}_k}^\bot  {\bf D}^H {\bf Y}^{(n)}_k{\boldsymbol a}^*(\tilde\theta) \|^2 d\tilde\theta \cdot \Delta\phi^2 + 2\Re\left( \int_{\tilde\theta-\theta_{\rm as}}^{{\tilde\theta+ \theta_{\rm as}}}
 {\boldsymbol a}^T(\tilde\theta) ({\bf Y}^{(n)}_k)^H {\bf P}_{{\bf B}_k}^\bot {\bf D}^H {\bf Y}^{(n)}_k  {\boldsymbol a}^*(\tilde\theta) d\tilde\theta
 \right) \Delta\phi \nonumber \\ & +
\int_{\tilde\theta_k-\theta_{\rm as}}^{{\tilde\theta_k + \theta_{\rm as}}} \|{\bf P}_{{\bf B}_k}^\bot  {\bf Y}^{(n)}_k{\boldsymbol a}^*(\tilde\theta)\|^2 d\tilde\theta, \label{equ2}
\end{align}
which is a parabola function with respect to $\Delta\phi$.  The above parabola function
achieves its minimum value when $\Delta\phi$ equals
\begin{align}
\Delta\phi^{(n+1)}_k (\tilde\theta) = \frac{  \Re\left(
  \int_{\tilde\theta_k-\theta_{\rm as}}^{{\tilde\theta_k + \theta_{\rm as}}} {\boldsymbol a}^T(\tilde\theta) ({\bf Y}^{(n)}_k)^H {\bf P}_{{\bf B}_k}^\bot {\bf D}^H {\bf Y}^{(n)}_k  {\boldsymbol a}^*(\tilde\theta)d\tilde\theta
 \right)  }{  \int_{\tilde\theta_k-\theta_{\rm as}}^{{\tilde\theta_k+ \theta_{\rm as}}} \| {\bf P}_{{\bf B}_k}^\bot {\bf D}^H  {\bf Y}^{(n)}_k{\boldsymbol a}^*(\tilde\theta) \|^2 d\tilde\theta }.
\end{align}

By substituting $\Delta\phi_k^{(n+1)}(\tilde\theta)$ for $\Delta\phi$ into (\ref{equ2}), the DOA estimation in the ($n+1$)-$\rm{th}$ iteration for the $k\rm{th}$ user can be expressed as
\begin{align}\label{equ3}
&\hat\theta_k^{(n+1)} =  \arg\min_{\tilde\theta_k} \frac{1}{ \int_{\tilde\theta_k-\theta_{\rm as}}^{{\tilde\theta_k + \theta_{\rm as}}} \| {\bf Y} {\boldsymbol a}^*(\tilde\theta) \|^2 d\tilde\theta } \times \nonumber \\ & \Bigg( \int_{\tilde\theta_k-\theta_{\rm as}}^{{\tilde\theta_k + \theta_{\rm as}}} \|{\bf P}_{{\bf B}_k}^\bot{\bf Y}^{(n)}_k{\boldsymbol a}^*(\tilde\theta)\|^2 d\tilde\theta  -\frac{  \Re\left( \int_{\tilde\theta_k -\theta_{\rm as}}^{{\tilde\theta_k + \theta_{\rm as}}}
 {\boldsymbol a}^T(\tilde\theta) ({\bf Y}^{(k)}_n)^H {\bf P}_{{\bf B}_k}^\bot {\bf D}^H {\bf Y}^{(n)}_k  {\boldsymbol a}^*(\tilde\theta) d\tilde\theta \right)^2
}{  \int_{\tilde\theta_k-\theta_{\rm as}}^{{\tilde\theta_k+ \theta_{\rm as}}} \| {\bf P}_{{\bf B}_k}^\bot{\bf D}^H {\bf Y}^{(n)}_k{\boldsymbol a}^*(\tilde\theta) \|^2 d\tilde\theta } \Bigg).
\end{align}

Accordingly, the CFO estimation can be updated as
\begin{align}
& \hat\phi^{(n+1)}_{k} = \hat\phi^{(n)}_{k} + \Delta\phi^{(n+1)}_k( \hat\theta^{(n+1)}_{k} ).
\end{align}

Note that optimization of (\ref{equ3}) can be efficiently solved by exploiting FFT operation. Specifically, let
${\boldsymbol y}(\tilde\theta) = {\bf Y}{\boldsymbol a}^*(\tilde\theta)$
represent the signal after the receive beamforming with steering vector ${\boldsymbol a}^*(\tilde\theta)$. We can rewrite (\ref{equ3}) into
\begin{align}\label{reviseequ1}
\hat\theta_k^{(n+1)} =  & \arg\max_{\tilde\theta_k}
 \frac{1}{ \int_{\tilde\theta_k-\theta_{\rm as}}^{{\tilde\theta_k + \theta_{\rm as}}} \| {\boldsymbol y}(\tilde\theta) \|^2 d\tilde\theta } \Bigg(
\int_{\tilde\theta_k-\theta_{\rm as}}^{{\tilde\theta_k + \theta_{\rm as}}} \|{\bf P}_{{\bf B}_k} {\bf E}^H(\hat\phi^{(n)}_k) {\boldsymbol y}(\tilde\theta) \|^2 d\tilde\theta \nonumber \\ & \kern 50pt +
\frac{  \Re\left( \int_{\tilde\theta_k -\theta_{\rm as}}^{{\tilde\theta_k + \theta_{\rm as}}}
 {\boldsymbol y}^H(\tilde\theta) {\bf E}(\hat\phi^{(n)}_k) {\bf P}_{{\bf B}_k} {\bf D}^H {\bf E}^H(\hat\phi^{(n)}_k) {\boldsymbol y}(\tilde\theta) d\tilde\theta \right)^2  }{  \int_{\tilde\theta_k-\theta_{\rm as}}^{{\tilde\theta_k+ \theta_{\rm as}}} \| {\bf D}^H {\boldsymbol y}(\tilde\theta) \|^2 d\tilde\theta -  \int_{\tilde\theta_k-\theta_{\rm as}}^{{\tilde\theta_k+ \theta_{\rm as}}} \| {\bf P}_{{\bf B}_k}{\bf D}^H {\bf E}^H(\hat\phi^{(n)}_k) {\boldsymbol y}(\tilde\theta) \|^2 d\tilde\theta  }
\Bigg).
\end{align}
\textcolor{black}{Note that ${\bf P}_{{\bf B}_k}$ has a dimension of $N\times N$, and thus directly computing (\ref{reviseequ1}) may suffer from a high computational burden.
To further reduce the computational complexity, }
we can express ${\bf P}_{{\bf B}_k}$ as ${\bf P}_{{\bf B}_k} = {\bf Q}_k{\bf Q}_k^H$ where ${\bf Q}_k \in\mathbb{C}^{N\times L}$ is a unitary matrix and has the same column space of ${\bf B}_k$, i.e., ${\rm Span}({\bf Q}_k)={\rm Span}({\bf B}_k)$.
 We further obtain
\begin{align}\label{equ40}
\hat\theta_k^{(n+1)}= & \arg\max_{\tilde\theta_k}
 \frac{1}{ \int_{\tilde\theta_k-\theta_{\rm as}}^{{\tilde\theta_k + \theta_{\rm as}}} \| {\boldsymbol y}(\tilde\theta) \|^2 d\tilde\theta } \Bigg(
\int_{\tilde\theta_k-\theta_{\rm as}}^{{\tilde\theta_k + \theta_{\rm as}}} \| {\bf Q}_k^H {\bf E}^H(\hat\phi^{(n)}_k) {\boldsymbol y}(\tilde\theta) \|^2 d\tilde\theta \nonumber \\ & \kern 50pt  +
\frac{  \Re\left( \int_{\tilde\theta_k -\theta_{\rm as}}^{{\tilde\theta_k + \theta_{\rm as}}}
 {\boldsymbol y}^H(\tilde\theta) {\bf E}(\hat\phi^{(n)}_k) {\bf Q}_k{\bf Q}_k^H {\bf D}^H {\bf E}^H(\hat\phi^{(n)}_k) {\boldsymbol y}(\tilde\theta) d\tilde\theta \right)^2  }{  \int_{\tilde\theta_k-\theta_{\rm as}}^{{\tilde\theta_k+ \theta_{\rm as}}} \| {\bf D}^H {\boldsymbol y}(\tilde\theta) \|^2 d\tilde\theta -  \int_{\tilde\theta_k-\theta_{\rm as}}^{{\tilde\theta_k+ \theta_{\rm as}}} \| {\bf Q}_k^H {\bf D}^H {\bf E}^H(\hat\phi^{(n)}_k) {\boldsymbol y}(\tilde\theta) \|^2 d\tilde\theta  }
\Bigg).
\end{align}

Let both the trial DOA $\tilde\theta_k$ and the DOA samples $\tilde\theta$ in the numerical integration be drawn from
$\mathcal{D}_{\rm FFT}=\{\mathcal{D}_1, \mathcal{D}_2, \cdots, \mathcal{D}_{M_{\rm FFT}}\}$ with $\mathcal{D}_i = \arccos \left( -\frac{2\pi(i-1)}{\chi M_{\rm FFT}} \right)$, where $M_{\rm FFT}$ represents the size of FFT.
There is ${\boldsymbol a}^*(\mathcal{D}_i) = [1, {\rm e}^{-{\bf j}2\pi (i-1)\frac{1}{M_{\rm FFT}}}, \cdots,$ ${\rm e}^{-{\bf j}2\pi (i-1)\frac{(M_{\rm FFT}-1)}{M_{\rm FFT}}}]^T$ and thus,
\begin{align}\label{equ20}
\Big[{\boldsymbol y}(\mathcal{D}_1),{\boldsymbol y}(\mathcal{D}_2), \cdots,{\boldsymbol y}(\mathcal{D}_{M_{\rm FFT}})\Big]
\end{align}
corresponds to the FFT conversion of ${\bf Y}$ along the spatial dimension. Denote the neighbor samples of $\tilde\theta_k$ by
\begin{align}
\mathcal{S}_{\tilde\theta_k} = \left\{ \mathcal{D}_i :  | \mathcal{D}_i - \tilde\theta_k|<\theta_{\rm as} \right\}.
\end{align}
The numerical solution of (\ref{equ40}) can be then expressed as
\begin{align}\label{equ41}
\hat\theta_k^{(n+1)}= & \arg\max_{\tilde\theta_k}
 \frac{1}{  \sum_{\tilde\theta\in \mathcal{S}_{\tilde\theta_k}}  \| {\boldsymbol y}(\tilde\theta) \|^2  } \Bigg(
\sum_{\tilde\theta\in \mathcal{S}_{\tilde\theta_k}}  \| {\bf Q}_k^H {\bf E}^H(\hat\phi^{(n)}_k) {\boldsymbol y}(\tilde\theta) \|^2 \nonumber \\ & \kern 50pt  +
\frac{  \Re\left( \sum_{\tilde\theta\in \mathcal{S}_{\tilde\theta_k}}
 {\boldsymbol y}^H(\tilde\theta) {\bf E}(\hat\phi^{(n)}_k) {\bf Q}_k{\bf Q}_k^H {\bf D}^H {\bf E}^H(\hat\phi^{(n)}_k) {\boldsymbol y}(\tilde\theta) \right)^2  }{  \sum_{\tilde\theta\in \mathcal{S}_{\tilde\theta_k}}  \| {\bf D}^H {\boldsymbol y}(\tilde\theta) \|^2  -  \| {\bf Q}_k^H {\bf D}^H {\bf E}^H(\hat\phi^{(n)}_k) {\boldsymbol y}(\tilde\theta) \|^2   }
\Bigg).
\end{align}
Note that the FFT operation in (\ref{equ20}) needs to be computed only once during the iterative estimation procedure for all users. Bearing in mind that the calculation of ${\boldsymbol y}(\mathcal{D}_i)$ dominates the computational burden of the grid search in (\ref{equ41}), the computational complexity can be greatly reduced by exploiting FFT in (\ref{equ20}).


\subsection{Multi-Branch Beamforming}

After the above joint CFO and DOA estimation, all steering vectors ${\boldsymbol a}^*(\tilde\theta)$ with $\tilde\theta\in (\hat\theta_k- \theta_{\rm as},\hat\theta_k+\theta_{\rm as})$ can be utilized as the receive beamforming vectors for the $k\rm{th}$ user, and the CFO estimate $\hat\phi_k$ can be used for compensation.
   With perfect estimation and infinite receive antennas, i.e., $\hat\phi_k = \phi_k$, $\hat\theta_k = \theta_k$, $M\to \infty$ and $\forall \tilde\theta\in (\theta_k- \theta_{\rm as},\theta_k+\theta_{\rm as})$,
  we can obtain the signals after beamforming corresponding to the training and data blocks in noise-free environment as follows:
\begin{align}
 & {\bf E}^H(\hat\phi_k){\bf Y}{\boldsymbol a}^*(\tilde\theta) = {\bf B}_k {\bf H}_k {\boldsymbol a}^*(\tilde\theta), \label{equ5}\\
 & \eta_i(\hat\phi_k)^*{\bf E}^H(\hat\phi_k){\bf Y}^{\rm d}_i{\boldsymbol a}^*(\tilde\theta) = \sqrt{N} {\bf F}^H \textrm{diag}({\bf s}_i^{(k)}){\bf F}_L {\bf H}_k {\boldsymbol a}^*(\tilde\theta).  \label{equ6}
\end{align}
The equivalent channel response ${\bf H}_k{\boldsymbol a}^*(\tilde\theta)$ can be estimated from (\ref{equ5}) by  the conventional least square (LS) algorithm, which can be further utilized for subsequent data detection in (\ref{equ6}).

However, in practice, the receiver has limited number of antennas and finite spatial resolution, and thus the MUI term in (\ref{equ8}) may not be negligible.
Hence, not all of the angles within the DOA region $(\theta_k-\theta_{\rm as}, \theta_k+\theta_{\rm as})$ are suitable as the beamforming direction for the $k\rm{th}$ user. For example,  once some users are close to each other in the spatial domain, beamforming towards
the angles on the edge of DOA region may suffer from some non-negligible
 MUI effect.
Therefore, instead of directly beamforming towards all angles within the DOA region $(\hat\theta_k-\theta_{\rm as}, \hat\theta_k+\theta_{\rm as})$, we  propose an alternative adaptive multi-branch beamforming scheme in this subsection.
\textcolor{black}{Note that after beamforming with ${\boldsymbol a}^*(\tilde\theta)$ and CFO compensation with $\hat\phi_k$ on the received signal matrix ${\bf Y}$, the magnitude of
\begin{align}\label{equ7}
\mathcal{C}_k(\tilde\theta,\hat\phi_k) = \frac{ \|{\bf P}_{{\bf B}_k}{\bf E}^H(\hat\phi_k){\bf Y}{\boldsymbol a}^*(\tilde\theta)\|^2} { \|{\bf P}_{{\bf B}_k}^\bot{\bf E}^H(\hat\phi_k){\bf Y}{\boldsymbol a}^*(\tilde\theta)\|^2 }
\end{align}
actually stands for the ratio of the expected signal power to the resultant noise power.
To some extent, we may consider the magnitude (\ref{equ7}) as the SNR for the $k\rm{th}$ user with beamforming vector of ${\boldsymbol a}^*(\tilde\theta)$.
Let us draw the trial DOAs  from set $\mathcal{D}_{\rm FFT}$.
When $\tilde\theta$ does not correspond to the $k\rm{th}$ user, the magnitude of $\mathcal{C}_k(\tilde\theta,\hat\phi_k)$ can be approximately given by $L/(N\!-\! L)$.
Then, we can define the set of qualified DOAs for the $k\rm{th}$ user as
\begin{align}
\vartheta^{(k)} =  \left\{ \mathcal{D}_i : \quad  \mathcal{C}_k(\mathcal{D}_i, \hat\phi_k) \ge t_h \cdot L/(N\!-\!L) \right\},
\end{align}
where $t_h$ is a pre-set threshold that can be larger than one. }
Here, the elements of $\vartheta^{(k)}$ are picked out from set $\mathcal{D}_{\rm FFT} $ in order to reuse the FFT operation on the received signal   (\ref{equ20}).
The basic idea of the proposed multi-branch beamforming is to use beamforming vectors
towards the qualified DOAs for each user. Mathematically, the corresponding receive beamforming matrix for the $k\rm{th}$ user can be obtained as
${\bf W}^{(k)} = \left[ {\boldsymbol a}^*(\vartheta^{(k)}_1), {\boldsymbol a}^*(\vartheta^{(k)}_2), \cdots, {\boldsymbol a}^*(\vartheta^{(k)}_{M_k})  \right] \in \mathbb{C}^{M\times M_k}$,
where $M_k$ is the cardinality of $\vartheta^{(k)}$ and $\vartheta^{(k)}_i$ is the $i\rm{th}$ element of $\vartheta^{(k)}$.

\subsection{Performance Analysis of  CFO Estimation}

In this subsection, we provide the theoretical performance analysis for the proposed CFO estimation.
For simplicity, we consider the case that BS has perfect knowledge of DOA information of each user,
and utilize $\left(\theta_k-\theta_{\rm as}, \theta_k+\theta_{\rm as} \right)$ as intergration interval in (\ref{costfunc}). Moreover, we consider the case that all the users have enough spatial separation between each other, and thus we can ignore the MUI effect during the analysis.
In this case, the CFO estimation for the $k\rm{th}$ user is equivalent to
\begin{align}
\hat\phi_k = \arg\min_{\tilde\phi} G_k(\tilde\phi),
\end{align}
where
$G_k(\tilde\phi) = \int_{\theta_k-\theta_{\rm as}}^{\theta_k+\theta_{\rm as}} {\boldsymbol a}^T(\tilde\theta){\bf Y}^H{\bf E}(\tilde\phi) {\bf P}_{{\bf B}_k}^\bot{\bf E}^H(\tilde\phi){\bf Y}{\boldsymbol a}^*(\tilde\theta) d\tilde\theta$.
The mean square error (MSE) of CFO estimation under the high SNR can be expressed as~\cite{Zhang15sub}
\begin{align}\label{equ10}
{\rm MSE}\{\hat\phi_k\} = \left.\frac{E[ ( \frac{\partial G_k(\tilde\phi)}{\partial \tilde\phi} )^2 ]}{ E[ \frac{\partial^2 G_k(\tilde\phi)}{\partial \tilde\phi^2}  ]^2 }\right|_{\tilde\phi = \phi_k}.
\end{align}

Bearing in mind of $\frac{\partial {\bf E}(\tilde\phi)}{\partial \tilde\phi} = {\bf D} {\bf E}(\tilde\phi)$, we obtain
\begin{align}
& \left. \frac{\partial G_k(\tilde\phi) }{\partial \tilde\phi} \right|_{\tilde\phi=\phi_k} \simeq  2\Re\left( \int_{\theta_k-\theta_{\rm as}}^{\theta_k+\theta_{\rm as}} {\boldsymbol a}^T(\tilde\theta){\bf N}^H  {\bf P}_{{\bf B}_k}^\bot{\bf D}^H{\bf B}_k {\bf H}_k{\boldsymbol a}^*(\tilde\theta) d\tilde\theta \right), \\
&  \left.  \frac{\partial^2 G_k(\tilde\phi) }{\partial \tilde\phi^2 } \right|_{\tilde\phi=\phi_k}  \simeq 2 \Big(  \int_{\theta_k-\theta_{\rm as}}^{\theta_k+\theta_{\rm as}}  {\boldsymbol a}^T(\tilde\theta) {\bf H}_k^H {\bf B}_k^H {\bf D} {\bf P}_{{\bf B}_k}^\bot{\bf D}^H {\bf B}_k {\bf H}_k {\boldsymbol a}^*(\tilde\theta)  d \tilde\theta \Big). \label{dev2nd}
\end{align}

Then, there is
\begin{align}\label{equ42}
& E\left[\left(\frac{\partial G_k(\tilde\phi) }{\partial \tilde\phi} \right)^2 \right] \Big|_{\tilde\phi=\phi_k}
=  2 E \left[\left | \int_{\theta_k-\theta_{\rm as}}^{\theta_k+\theta_{\rm as}} {\boldsymbol a}^T(\tilde\theta){\bf N}^H {\bf P}_{{\bf B}_k}^\bot{\bf D}^H{\bf B}_k {\bf H}_k{\boldsymbol a}^*(\tilde\theta) d\tilde\theta  \right|^2\right] \nonumber \\
= & 2 E\left[  \int_{\theta_k-\theta_{\rm as}}^{\theta_k+\theta_{\rm as}}\int_{\theta_k-\theta_{\rm as}}^{\theta_k+\theta_{\rm as}}
{\boldsymbol a}^T(\tilde\theta){\bf N}^H {\bf P}_{{\bf B}_k}^\bot{\bf D}^H{\bf B}_k {\bf H}_k{\boldsymbol a}^*(\tilde\theta)
{\boldsymbol a}^T(\tilde\alpha) {\bf H}_k^H  {\bf B}_k^H {\bf D} {\bf P}_{{\bf B}_k}^\bot {\bf N} {\boldsymbol a}^*(\tilde\alpha)
d\tilde\theta d\tilde\alpha
  \right] \nonumber \\
  = & 2\sigma_n^2
   \int_{\theta_k-\theta_{\rm as}}^{\theta_k+\theta_{\rm as}}\int_{\theta_k-\theta_{\rm as}}^{\theta_k+\theta_{\rm as}}
   \textrm{Tr}\Big( {\bf H}_k^H  {\bf B}_k^H {\bf D} {\bf P}_{{\bf B}_k}^\bot  {\bf D}^H{\bf B}_k {\bf H}_k{\boldsymbol a}^*(\tilde\theta)
   {\boldsymbol a}^T(\tilde\theta) {\boldsymbol a}^*(\tilde\alpha) {\boldsymbol a}^T(\tilde\alpha) \Big)   d\tilde\theta d\tilde\alpha.
\end{align}

Define
${\bf R}_k  =    \int_{\theta_k-\theta_{\rm as}}^{\theta_k+\theta_{\rm as}}  \frac{1}{2\theta_{\rm as }}{\boldsymbol a}^*(\tilde\theta)
   {\boldsymbol a}^T(\tilde\theta) d\tilde\theta$
with its ($i,j$)-$\rm{th}$ entry being
${\bf R}_k(i,j) = \frac{1}{2\theta_{\rm as}}\int_{\theta_k-\theta_{\rm as}}^{\theta_k+\theta_{\rm as}} {\rm e}^{{\bf j}\chi (i-j)\cos\tilde\theta}   d \tilde\theta$.
We can further simplify (\ref{equ42}) as
\begin{align}
 E\left[\left(\frac{\partial G_k(\tilde\phi) }{\partial \tilde\phi} \right)^2 \right] \Big|_{\tilde\phi=\phi_k} = & 8\sigma_n^2 \theta_{\rm as}^2\cdot \textrm{Tr}\Big( {\bf H}_k^H {\bf B}_k^H {\bf D}{\bf P}_{{\bf B}_k}^\bot {\bf D}^H {\bf B}_k {\bf H}_k {\bf R}_k {\bf R}_k  \Big) \nonumber \\
  = &   8\sigma_n^2 \theta_{\rm as}^2\cdot \textrm{Tr}\Big(  {\bf B}_k^H {\bf D}{\bf P}_{{\bf B}_k}^\bot {\bf D}^H {\bf B}_k {\bf H}_k {\bf R}_k {\bf R}_k {\bf H}_k^H \Big) \nonumber \\
   \simeq & \textcolor{black}{ 8\sigma_n^2 \theta_{\rm as}^2 \cdot \textrm{Tr}\big({\bf R}_k^3 \big) \cdot \|  {\bf P}_{{\bf B}_k}^\bot {\bf D}^H {\bf B}_k {\boldsymbol\Sigma}_h\|^2 , } \label{equ43}
\end{align}
\textcolor{black}{where ${\boldsymbol\Sigma}_h = \textrm{diag}( \sigma_{h,1}, \sigma_{h,2}, \cdots, \sigma_{h,L} )$ corresponds to the channel PDP.
The derivation of (\ref{equ43}) arises from the fact that $E\big[  ({\bf h}_l^{(k)})^*  ( {\bf h}_l^{(k)} )^T   \big]  =   \sigma_{h,l}^2 \cdot   E\big[ \sum_{p=1}^P \frac{1}{P}
  {\boldsymbol a}^*(\theta_{l,p,k})    {\boldsymbol a}^T(\theta_{l,p,k})    \big]) =   \sigma_{h,l}^2 \cdot     {\bf R}_k$. }

For the second order derivative in (\ref{dev2nd}), we can obtain the following approximation:
\begin{align}\label{equ11}
\left.  \frac{\partial^2 G_k(\tilde\phi) }{\partial \tilde\phi^2 } \right|_{\tilde\phi=\phi_k}  & \simeq 2\int_{\theta_k-\theta_{\rm as}}^{\theta_k+\theta_{\rm as}}  {\boldsymbol a}^T(\tilde\theta){\bf H}_k^H {\bf B}_k^H {\bf D} {\bf P}_{{\bf B}_k}^\bot{\bf D}^H {\bf B}_k{\bf H}_k {\boldsymbol a}^*(\tilde\theta) d\tilde\theta \nonumber \\  \simeq &
 2 \int_{\theta_k-\theta_{\rm as}}^{\theta_k+\theta_{\rm as}}
 {\boldsymbol a}^T(\tilde\theta) {\bf R}_k {\boldsymbol a}^*(\tilde\theta)  d\tilde\theta \cdot
 \textcolor{black}{
 \| {\bf P}_{{\bf B}_k}^\bot{\bf D}^H {\bf B}_k {\bf\Sigma}_h \|^2 } \nonumber \\
= &  4\theta_{\rm as}  \cdot\| {\bf R}_k \|^2  \cdot  \textcolor{black}{
 \| {\bf P}_{{\bf B}_k}^\bot{\bf D}^H {\bf B}_k {\bf\Sigma}_h \|^2 }.
 \end{align}

By substituting (\ref{equ43}) and (\ref{equ11}) into (\ref{equ10}), the CFO estimation MSE for the $k\rm{th}$ user can be expressed as:
\begin{align}\label{MSEhatphik}
\textrm{MSE}\{\hat\phi_k\} = \frac{   \sigma_n^2  \cdot  \textrm{Tr}\big({\bf R}_k^3 \big)  }{ 2\cdot \| {\bf R}_k\|^4 \cdot \textcolor{black}{ \| {\bf P}_{{\bf B}_k}^\bot{\bf D}^H {\bf B}_k  {\bf\Sigma}_h \|^2 } }.
\end{align}

\textcolor{black}{Let $\sigma_s^2$ denote the transmit signal power from each user. With  sufficiently large $M$ and $N$, the asymptotic CFO estimation MSE for the $k\rm{th}$ user can be expressed as
\begin{align}\label{MSEasym}
\textrm{MSE}_{asym}\{\hat\phi_k\} =  \frac{ 3 \sigma_n^2 / \sigma_s^2   }{ 2 \pi^2 M N  },
\end{align}
indicating that the estimation MSE decreases as $1/M$. In other words, for a fixed desired MSE, the required transmit power of each user decreases as $1/M$. The detailed derivation of (\ref{MSEasym}) can be found in Appendix A.
}


\section{Improved Scheme with User Grouping}

Due to user mobility, there exists the possibility that some users may have very close DOAs, or even overlapped DOAs.
Meanwhile, the massive BS cannot have infinite number of antennas in practice, which indicates non-ideal spatial resolution.
Thus, the proposed scheme in Section III may not work properly once the signal from multiple users cannot be separated well from the spatial domain, which can substantially deteriorate the CFO estimation and the subsequent data detection.
Considering the random distribution of multiple users, we here propose an improved frequency synchronization scheme with the aid of user grouping, whose details are  described as follows.

Here we should note that, the proposed scheme in Section III is originally designed for users with sufficient spatial separation; Nevertheless,
 for the users with a certain MUI effect, the previously obtained CFO estimate and qualified DOAs can still serve as valid coarse estimation, which can be employed
 to facilitate the following user grouping procedure.  Specifically, according to the discussions on (\ref{equ7}),
 the resultant SNR with beamforming matrix  ${\bf W}^{(k)}$ can be expressed as $\rho_k = \sum_{\tilde\theta\in \vartheta^{(k)} } \mathcal{C}_k(\tilde\theta,\hat\phi_k)$.
Moreover, the total received signal power at the BS can be approximately expressed as $\hat\sigma_s^2 = \|{\bf Y}\|^2$. The noise power can be estimated as
$\hat\sigma_n^2 = \min\limits_{\tilde\theta\in \mathcal{D}_{\rm FFT}} \|{\bf Y}{\boldsymbol a}^*(\tilde\theta)\|^2$, where the trial beamforming vector ${\boldsymbol a}^*(\tilde\theta)$ aims at suppressing the signals from all users and thus the resultant ${\bf Y}{\boldsymbol a}^*(\tilde\theta)$ mainly corresponds to the noise term.
Then, the expected average SNR for each user can be evaluated as $\rho_{\rm ex} = \frac{\hat\sigma_s^2/K}{\hat\sigma_n^2}$.
Once the estimated SNR of the $k\rm{th}$ user is below a threshold, say $\rho_k < \rho_{\rm th} \cdot \rho_{\rm ex}$ where $\rho_{\rm th}$ is a pre-set value,
there may exist non-negligible MUI from other users. In this case, we denote the $k\rm{th}$ user as a \textit{critical} user; otherwise, we denote it as a \textit{non-critical} user.

The grouping strategy in angle domain includes the following two steps:
First, we gather multiple critical users with similar DOAs into one group; namely, in one group the distance of the qualified DOAs between any user and the rest users should be less than a certain guard interval $\mathcal{G}$. Mathematically, denote the user index set of the $g\rm{th}$ group  by $\mathcal{U}^{(g)} = \{c^{(g)}_1,\cdots, c^{(g)}_{K_g}\}$,
where $K_g$ is the number of the users in this group. Then, there should be
$\min\limits_{j\ne i}   {\rm dist}( \vartheta^{(c^{(g)}_i)}, \vartheta^{(c^{(g)}_j)} ) < \mathcal{G}$ for any $i=1,2,\cdots,K_g$, where ${\rm dist}( \mathcal{U}_1, \mathcal{U}_2 )  \triangleq \max |c_1-c_2|$, $\forall c_1\in \mathcal{U}_1$ and $\forall c_2\in \mathcal{U}_2$.
Second, for each group, we also take in the non-critical users whose minimal distance from all critical users within this group is smaller than the guard interval $\mathcal{G}$.

If one user is not grouped with all the other users, then this user is well separated from the others in the spatial domain, and the previously designed CFO estimation, beamforming, as well as data detection can be applied immediately for this user.
On the contrary, for the user grouped with some other users, the non-negligible MUI would happen among the users in the same group. In this case, the CFO estimation and data detection should be jointly carried out for the multiple users in the same group.

Therefore, we here develop a joint CFO estimation and data detection scheme for the grouped users.  Without loss of generality, we consider that the first $\kappa$ users stay in one group, i.e., $\mathcal{U}^{(1)}=\{1,2,\cdots,\kappa\}$, and illustrate the the CFO estimation and data detection for this group.  Similar approach can be readily extended to other groups.

\subsection{CFO Estimation}

Suppose the DOA range for the considered group is [$\theta_{\min}, \theta_{\max}$], where $\theta_{\min}$ and $\theta_{\max}$ are the minimum and maximum qualified DOAs from all $\kappa$ users, respectively, i.e., $\theta_{\min} = \min  \Big( \bigcup\nolimits_{k=1,\cdots,\kappa}\vartheta^{(k)} \Big)$ and $\theta_{\max} = \max \Big( \bigcup\nolimits_{k=1,\cdots,\kappa}\vartheta^{(k)} \Big)$.
Let us then pick up the following $Q=\kappa L$ uniformly distributed DOAs within  $[\theta_{\min}, \theta_{\max}]$ to formulate the  beamforming vectors for this group:
$\psi_{i} = \theta_{\min}  + (i-1)\frac{\theta_{\max}-\theta_{\min} }{Q-1}$, $i=1,2,\cdots,Q$.
The corresponding beamforming matrix  can be obtained as ${\bf W}_{\rm Grp} = \big[ {\boldsymbol a}(\psi_{1}), {\boldsymbol a}(\psi_{2}), \cdots, {\boldsymbol a}(\psi_{Q})  \big]^*$,
and the beamforming is  performed as
\begin{align}\label{equ21}
{\bf Y}_{\rm Grp} = {\bf Y}{\bf W}_{\rm Grp}.
\end{align}
Note that the user outside this group should have enough spatial separation from the users in this group, as defined by the grouping strategy, which says that the beamforming matrix ${\bf W}_{\rm Grp}$ could substantially mitigate the MUI from the users outside the group.
Then, we can approximately consider that ${\bf Y}_{\rm Grp}$ is composed of only the $\kappa$ users in this group.

Define
${\mathbb B}({\boldsymbol\phi}) = \Big[ {\bf E}(\phi_1) {\bf B}_1, {\bf E}(\phi_2) {\bf B}_2, \cdots, {\bf E}(\phi_{\kappa}) {\bf B}_{\kappa} \Big] \in \mathbb{C}^{N\times \kappa L}$
where ${\boldsymbol\phi} = [\phi_1, \phi_2, \cdots, \phi_{\kappa}  ]^T$.
Denote
${\bf H}_{{\rm eq}}^{(k)} =  {\bf H}_k {\bf W}_{\rm Grp}  \in \mathbb{C}^{L\times Q}$ and
 ${\bf H}_{\rm eq} = \Big[ ({\bf H}_{{\rm eq}}^{(1)})^T, ({\bf H}_{{\rm eq}}^{(2)})^T, \cdots, ({\bf H}_{{\rm eq}}^\kappa)^T  \Big]^T \in \mathbb{C}^{\kappa L\times Q}$,
and rewrite ${\bf Y}_{\rm Grp}$ as
\begin{align}\label{equ18}
{\bf Y}_{\rm Grp} = {\mathbb B}({\boldsymbol\phi}) {\bf H}_{\rm eq} + {\bf N}_{\rm Grp},
\end{align}
where ${\bf N}_{\rm Grp} = {\bf N} {\bf W}_{\rm Grp} + \sum_{q=\kappa+1}^K  {\bf E}(\phi_q) {\bf B}_q {\bf H}_q  {\bf W}_{\rm Grp}$ denotes the corresponding noise plus the mitigated MUI terms from the  users outside the group.

Let $\tilde{\boldsymbol\phi} = [ \tilde\phi_1, \tilde\phi_2,\cdots, \tilde\phi_{\kappa} ]^T$ represent the trial value of ${\boldsymbol \phi}$.
 According to (\ref{equ18}), the LS joint CFO estimation of the $\kappa$ users in this group can be given by
 \begin{align}\label{equ13}
 \hat{\boldsymbol\phi}_{\rm Grp} = \arg \min_{\tilde{\boldsymbol\phi}} C_{\rm Grp}( \tilde{\boldsymbol\phi} )
 \end{align}
where
the cost function is $C_{\rm Grp}( \tilde{\boldsymbol\phi} ) = \big\| {\bf P}^\bot_{{\mathbb B}}(\tilde{\boldsymbol\phi}) {\bf R}_{\rm Grp} \big\|^2$, ${\bf R}_{\rm Grp} = {\bf Y}_{\rm Grp}{\bf Y}_{\rm Grp}^H$,
 ${\bf P}^\bot_{{\mathbb B}}(\tilde{\boldsymbol\phi}) = {\bf I}_N - {\bf P}_{{\mathbb B}}(\tilde{\boldsymbol\phi})$ and ${\bf P}_{{\mathbb B}}(\tilde{\boldsymbol\phi}) = \mathbb{B}({\boldsymbol\phi}) (\mathbb{B}^H({\boldsymbol\phi})\mathbb{B}({\boldsymbol\phi}))^{-1} \mathbb{B}^H({\boldsymbol\phi})$ is the orthogonal projection on the column space of $\mathbb{B}({\boldsymbol\phi})$.

Unfortunately, directly solving (\ref{equ13}) requires multi-dimensional search, which is computationally inefficient. Note that although the MUI effect within the group cannot be easily mitigated from the spatial domain, previous proposed CFO estimate  (\ref{equ1}), i.e., $\hat\phi_k$, may still be able to serve as a valid coarse estimation. \textcolor{black}{
In the following, we propose to approximate the cost function $C_{\rm Grp}(\tilde{\boldsymbol \phi})$ as a high dimensional parabola in the neighborhood of coarse estimation $\hat{\boldsymbol \phi}= [\hat\phi_1, \hat\phi_2,\cdots, \hat\phi_{\kappa} ]^T$, and the minimization of (\ref{equ13}) can be then solved in a more efficient way. }
Specifically, there is
\begin{align}
{\bf \Pi}(\tilde{\boldsymbol\phi}) =  {\mathbb B}(\tilde{\boldsymbol\phi}){\mathbb B}(\tilde{\boldsymbol\phi})^H  = \sum_{k=1}^{\kappa} {\bf E}(\tilde\phi_k){\bf B}_{k} {\bf B}_{k}^H {\bf E}(\tilde\phi_k)^H.
\end{align}
We have the approximation
${\bf E}(\tilde\phi_k){\bf B}_{k} = {\bf E}(\hat\phi_{k}){\bf B}_{k} +\Delta{\bf B}_{k}$,
where $\Delta{\bf B}_k  = {\bf D} {\bf B}_k \Delta\phi_k + \frac{1}{2}{\bf D}^2 {\bf B}_k \Delta\phi_k^2$ and $\Delta\phi_k = \tilde\phi_k - \hat\phi_k$.
We can further obtain
\begin{align}
{\bf\Pi}(\tilde{\boldsymbol\phi}) = {\bf\Pi}(\hat{\boldsymbol\phi}) + \Delta{\bf \Pi},
\end{align}
where
\begin{align}
\Delta{\bf \Pi} = & \sum_{k=1}^\kappa {\bf B}_k \Delta{\bf B}_k^H + \Delta{\bf B}_k {\bf B}_k^H + \Delta{\bf B}_k\Delta{\bf B}_k^H  \nonumber \\
= & \sum_{k=1}^\kappa ( {\bf B}_k {\bf B}_k^H {\bf D}^H + {\bf D}{\bf B}_k {\bf B}_k^H ) \Delta\phi_k +
\frac{1}{2}( {\bf B}_k {\bf B}_k^H {\bf D}^{2H} + {\bf D}^2{\bf B}_k {\bf B}_k^H + 2{\bf D}{\bf B}_k{\bf B}_k^H {\bf D}^H  ) \Delta\phi_k^2.  \label{equ14}
\end{align}

Performing eigenvalue decomposition on ${\bf\Pi}(\hat{\boldsymbol\phi})$ gives
\begin{align}
{\bf\Pi}(\hat{\boldsymbol\phi})  = \big[{\bf U}_{{\rm s}}, {\bf U}_{{\rm n}} \big] {\bf\Sigma}_{{\bf\Pi}} \big[{\bf U}_{{\rm s}}, {\bf U}_{{\rm n}} \big]  ^H,
\end{align}
where ${\bf U}_{\rm s} \in \mathbb{C}^{N\times \kappa L} $ and ${\bf U}_{\rm n} \in \mathbb{C}^{N\times (N-\kappa L)}$ correspond to the signal and noise subspace, respectively. Then, there is ${\bf P}^\bot_{{\mathbb B}}(\hat{\boldsymbol\phi}) = {\bf U}_{\rm n} {\bf U}_{\rm n}^H$.
Denote the $i\rm{th}$ eigenvalue (in ascending order) and the corresponding eigenvector of ${\bf\Pi}(\hat{\boldsymbol\phi})$ by $\lambda_i$ and ${\boldsymbol e}_i$, $i=1,2,\cdots, L$, respectively.
Then, the $i\rm{th}$ eigenvector of ${\bf\Pi}(\tilde{\boldsymbol\phi})$ can be expressed as ${\boldsymbol e}_i + \Delta {\boldsymbol e}_i$, where the perturbation term is given by~\cite{Wilkinson}
\begin{align}\label{equ15}
\Delta {\boldsymbol e}_i = \sum_{j\ne i} {\boldsymbol e}_j  \frac{{\boldsymbol e}_j^H   \Delta{\bf\Pi} {\boldsymbol e}_i}{\lambda_i - \lambda_j}.
\end{align}

Bearing in mind that
\begin{align}
{\bf P}^\bot_{{\mathbb B}}(\tilde{\boldsymbol\phi})  = \sum_{i=1}^{N-\kappa L}  ({\boldsymbol e}_i + \Delta {\boldsymbol e}_i) ({\boldsymbol e}_i + \Delta {\boldsymbol e}_i)^H,
\end{align}
we further obtain the approximation
\begin{align}\label{equ16}
C_{\rm Grp}( \tilde{\boldsymbol\phi} )  \simeq  &
 \sum_{i=1}^{N-\kappa L} ({\boldsymbol e}_i + \Delta {\boldsymbol e}_i)^H {\bf R}_{\rm Grp} ({\boldsymbol e}_i + \Delta {\boldsymbol e}_i) \\  = &  \textrm{Tr}( {\bf U}^H_{\rm n} {\bf R}_{\rm Grp} {\bf U}_{\rm n} )   +  \sum_{i=1}^{N-\kappa L}    {\boldsymbol e}_i^H {\bf R}_{\rm Grp} \Delta{\boldsymbol e}_i +  \Delta{\boldsymbol e}_i^H {\bf R}_{\rm Grp} {\boldsymbol e}_i +  \Delta{\boldsymbol e}_i^H {\bf R}_{\rm Grp} \Delta{\boldsymbol e}_i.
\end{align}

By substituting (\ref{equ14}) into (\ref{equ15}), we obtain
\begin{align}
\Delta{\boldsymbol e}_i =  \sum_{k=1}^\kappa    {\boldsymbol \alpha}_{k,i} \Delta\phi_k + {\boldsymbol \beta}_{k,i} \Delta\phi_k^2,
\end{align}
where
\begin{align}
& {\boldsymbol \alpha}_{k,i} = \sum_{j\ne i} {\boldsymbol e}_j  \frac{{\boldsymbol e}_j^H   \left( {\bf B}_k {\bf B}_k^H {\bf D}^H + {\bf D}{\bf B}_k {\bf B}_k^H \right){\boldsymbol e}_i}{\lambda_i - \lambda_j} = \sum_{j=N-\kappa L+1}^N {\boldsymbol e}_j  \frac{{\boldsymbol e}_j^H  {\bf B}_k {\bf B}_k^H {\bf D}^H {\boldsymbol e}_i}{- \lambda_j}, \\
& {\boldsymbol \beta}_{k,i} = \frac{1}{2}\sum_{j\ne i} {\boldsymbol e}_j  \frac{{\boldsymbol e}_j^H   \left( {\bf B}_k {\bf B}_k^H {\bf D}^{2H} + {\bf D}^2{\bf B}_k {\bf B}_k^H + 2{\bf D}{\bf B}_k{\bf B}_k^H {\bf D}^H \right){\boldsymbol e}_i}{\lambda_i - \lambda_j}.
\end{align}

With (\ref{equ16}), we may rewrite the cost function as
\begin{align}
 C_{\rm Grp}( \tilde{\boldsymbol\phi} )  \simeq  &  \textrm{Tr}( {\bf U}^H_{\rm n} {\bf R}_{\rm Grp} {\bf U}_{\rm n} ) + \sum_{i=1}^{N-\kappa L}  \Re\left( {\boldsymbol e}_i^H {\bf R}_{\rm Grp} \left(\sum_{k=1}^K    {\boldsymbol \alpha}_{k,i} \Delta\phi_k + {\boldsymbol \beta}_{k,i} \Delta\phi_k^2\right) \right)  \nonumber \\ & \kern 50pt +  \left(\sum_{k=1}^K    {\boldsymbol \alpha}_{k,i} \Delta\phi_k \right)^H {\bf R}_{\rm Grp} \left(\sum_{k=1}^K    {\boldsymbol \alpha}_{k,i} \Delta\phi_k \right).
\end{align}

Letting the first derivative of  $C_{\rm Grp}( \tilde{\boldsymbol\phi} ) $ with respect to $\Delta\phi_k$, $k=1,2,\cdots,\kappa,$ be zero yields
\begin{align}
& 2\Re\left( \sum_{i=1}^{N-KL} {\boldsymbol e}_i^H {\bf R}_{\rm Grp} {\boldsymbol\alpha}_{k,i} \right) +
4\Re\left( \sum_{i=1}^{N-\kappa L} {\boldsymbol e}_i^H {\bf R}_{\rm Grp} {\boldsymbol\beta}_{k,i} \right)\Delta\phi_k \nonumber \\ & \kern 140pt +  2\sum_{q=1}^K \Re\left( \sum_{i=1}^{N-\kappa L}  {\boldsymbol\alpha}_{k,i}^H {\bf R}_{\rm Grp}      {\boldsymbol \alpha}_{q,i} \right) \Delta\phi_q = 0. \label{equ17}
\end{align}

Denote
\begin{align}
& \mathbb{B}_k = -{\bf U}_{\rm s} \textrm{diag}(\frac{1}{{\boldsymbol\lambda}_{\rm s}}) {\bf U}_{\rm s}^H {\bf B}_k{\bf B}_k^H {\bf D}^H {\bf U}_{\rm n}, \\
& \mathbb{D}_k = -{\bf U}_{\rm s} \textrm{diag}(\frac{1}{{\boldsymbol\lambda}_{\rm s}}) {\bf U}_{\rm s}^H ( {\bf B}_k{\bf B}_k^H {\bf D}^{2H} + 2{\bf D}{\bf B}_k{\bf B}_k^H {\bf D}^H ) {\bf U}_{\rm n}.
\end{align}

Then, (\ref{equ17}) can be expressed as
\begin{align}\label{equ30}
& \Re\big( \textrm{Tr}( {\bf U}_{\rm n}^H {\bf R}_{\rm Grp} \mathbb{B}_k ) \big) + \Re\big( \textrm{Tr}( {\bf U}_{\rm n}^H {\bf R}_{\rm Grp}  \mathbb{D}_k ) \big)\Delta\phi_k + \sum_{q=1}^\kappa \Re\big( \textrm{Tr}( \mathbb{B}_k^H {\bf R}_{\rm Grp} \mathbb{B}_q ) \big)\Delta\phi_q = 0, \nonumber \\ & \kern 300pt k=1,2,\cdots,\kappa,
\end{align}
which would provide a linear estimation for  $\Delta\phi_k$.
Hence, the CFO estimation for the $\kappa$ users can be updated as
$ \hat{\boldsymbol\phi}_{\rm Grp} = [ \hat\phi_{1,{\rm Grp}}, \hat\phi_{2,{\rm Grp}}, \cdots, \hat\phi_{\kappa,{\rm Grp}}  ]^T$  with $\hat\phi_{k,{\rm Grp}} = \hat\phi_k + \Delta\phi_k$, $k=1,2,\cdots,\kappa$.


\subsection{Data Detection}

After the CFO estimation, the equivalent channel response matrix can be obtained from (\ref{equ18}) via the LS algorithm:
$\hat{\bf H}_{\rm eq} = \mathbb{B}^\dag(\hat{\boldsymbol\phi}_{\rm Grp}) {\bf Y}_{\rm Grp}$.
Similar to (\ref{equ21}), the beamforming can  also be performed for the data blocks as:
\begin{align}\label{equ19}
{\bf Y}_{i,{\rm Grp}}^{\rm d} = {\bf Y}_{i}^{\rm d} {\bf W}_{\rm Grp} =  \sqrt{N} \sum_{k=1}^\kappa \eta_i(\phi_k) {\bf E}(\phi_k) {\bf F}^H \textrm{diag}({\bf s}_i^{(k)}){\bf F}_L {\bf H}_{\rm eq}^{(k)} + {\bf N}_{i,{\rm Grp}}^{\rm d},
\end{align}
where  ${\bf N}_{i,{\rm Grp}}^{\rm d} \in \mathbb{C}^{N\times Q}$ denotes the corresponding noise plus the interference from the users outside the group.
We rewrite the equivalent channel matrix for the $k\rm{th}$ user in (\ref{equ19}) as
$ {\bf H}_{{\rm eq}}^{(k)} = \big[{\bf h}_{1,{\rm eq}}^{(k)}, {\bf h}_{2,{\rm eq}}^{(k)}, \cdots, {\bf h}_{Q,{\rm eq}}^{(k)}\big]$.
Let $\mathcal{H}_{q, {\rm eq}}^{(k)} \in\mathbb{C}^{N\times N}$ stand for the circular convolution matrix corresponding to the equivalent channel response ${\bf h}^{(k)}_{q,{\rm eq}}$, i.e.,
\begin{align}\label{equ22}
\mathcal{H}_{q, {\rm eq}}^{(k)} = \sqrt{N} {\bf F} \textrm{diag}( {\bf F}_L {\bf h}^{(k)}_{q,{\rm eq}}) {\bf F}^H.
\end{align}
The vectorization of (\ref{equ19}) is
\begin{align}
\textrm{vec}({\bf Y}_{i,{\rm Grp}}^{\rm d}) =&  \sum_{k=1}^\kappa \eta_i(\phi_k) ({\bf I}_Q\otimes {\bf E}(\phi_k)) \mathcal{H}_{\rm eq}^{(k)} {\bf F}^H {\bf s}_i^{(k)} + \textrm{vec}({\bf N}_{i,{\rm Grp}}^{\rm d} ) \\
= & \mathcal{H}_{\rm eq} ({\boldsymbol\eta}_i\otimes {\bf F}^H) {\bf s}_i + \textrm{vec}({\bf N}_{i,{\rm Grp}}^{\rm d} ),
\end{align}
where
\begin{align}
& {\bf s}_i = \big[ ({\bf s}_i^{(1)})^T, ({\bf s}_i^{(2)})^T, \cdots, ({\bf s}_i^{(\kappa)})^T  \big]^T, \\
& {\boldsymbol\eta}_i({\boldsymbol\phi}) = \textrm{diag}\big( \eta_i(\phi_1), \eta_i(\phi_2),\cdots, \eta_i(\phi_\kappa) \big), \\
& \mathcal{H}_{\rm eq}^{(k)} = ({\bf I}_Q\otimes {\bf E}(\phi_k)) \big[ (\mathcal{H}_{1, {\rm eq}}^{(k)})^T, (\mathcal{H}_{2, {\rm eq}}^{(k)} )^T, \cdots, (\mathcal{H}_{Q, {\rm eq}}^{(k)})^T \big ]^T \in\mathbb{C}^{QN\times N},  \label{equ23}\\
& \mathcal{H}_{\rm eq}= [ \mathcal{H}_{{\rm eq}}^{(1)}, \mathcal{H}_{{\rm eq}}^{(2)}, \cdots, \mathcal{H}_{{\rm eq}}^{(\kappa)}] \in\mathbb{C}^{ QN\times \kappa N}. \label{equ24}
\end{align}

Then, the ZF data detection for the $\kappa$ users can be performed as
$\hat{\bf s}_i = ({\boldsymbol\eta}_i^*({\boldsymbol\phi})\otimes {\bf F}) \mathcal{H}_{\rm eq}^{\dag} \textrm{vec}({\bf Y}^{\rm d}_{i,{\rm Grp}})$.
In practice, ${\boldsymbol\phi}$ can be replaced by $\hat{\boldsymbol\phi}_{\rm Grp}$, while $\mathcal{H}_{\rm eq}$ can be constructed from both $\hat{\bf H}_{\rm eq}$ and $\hat{\boldsymbol\phi}_{\rm Grp}$ in the way as (\ref{equ22}), (\ref{equ23}) and (\ref{equ24}).

\section{Simulations}

In this section, we demonstrate the effectiveness of the proposed frequency synchronization scheme through numerical examples.  16-QAM constellation is adopted for data transmission. In the proposed scheme, the training symbols are randomly drawn from the QPSK constellations. \textcolor{black}{The SNR is defined as $\sigma_s^2/\sigma_n^2$.}
The channel length is set as $L=10$ and the length of CP is taken as $N_{\rm cp}=L\!-\!1=9$.
Unless otherwise stated, the total number of subcarriers is taken as $N=64$, while we consider $M=128$ and $d=\lambda/2$. The channel vectors of different users are formulated according to (\ref{equ25}) with uniform PDP, i.e., $\sigma_{h,l}^2=1/L$.
 Both angular spreads of $\theta_{\rm as}=5^\circ$ and $\theta_{\rm as}=10^\circ$ are considered, while the guard interval is taken as $\mathcal{G}=4\theta_{\rm as}+5^\circ$.
Unless otherwise stated, we assume perfect knowledge about angular spread is available at BS. 
 The FFT size is set as $M_{\rm FFT}=2M$.
The normalized CFO is randomly generated from $-\phi_{\max}$ to $\phi_{\max}$.
The MSE of the normalized CFO is adopted as the figure of merit.
The thresholds are taken as $t_h=10$ and $\rho_{\rm th}=2/3$. The proposed scheme and the improved scheme with user grouping are labelled as `FS-BEAM' and `FS-BEAM-UG'.
For comparison, we also include the results of the existing robust multi-CFO estimation scheme with iterative interference cancellation of~\cite{Tsai13}, referred to as `RMCE-IIC'. Specifically, the ZC sequence in RMCE-IIC is adopted as $\exp({\bf j}\pi i^2/N)$, $i=0,1,\cdots,N-1$. The circular-shift set follows the constraint of~\cite[eq. (42)]{Tsai13}. We consider that three iterations of iterative interference cancellation is adopted in RMCE-IIC.

\begin{figure}[t]
\begin{center}
\includegraphics[width=100mm]{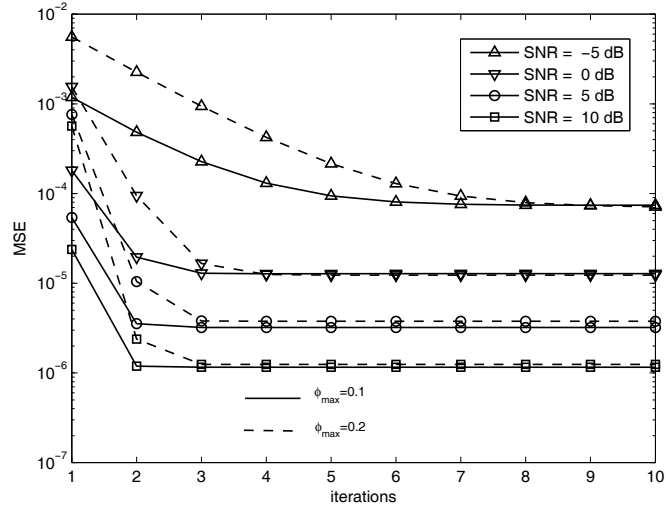}
\end{center}
\caption{
Convergence procedure of the CFO estimation performance of the proposed FS-BEAM scheme ($K=4$, $\theta_{\rm as}=5^\circ$).
}
\end{figure}

\begin{figure}[t]
\begin{center}
\includegraphics[width=120mm]{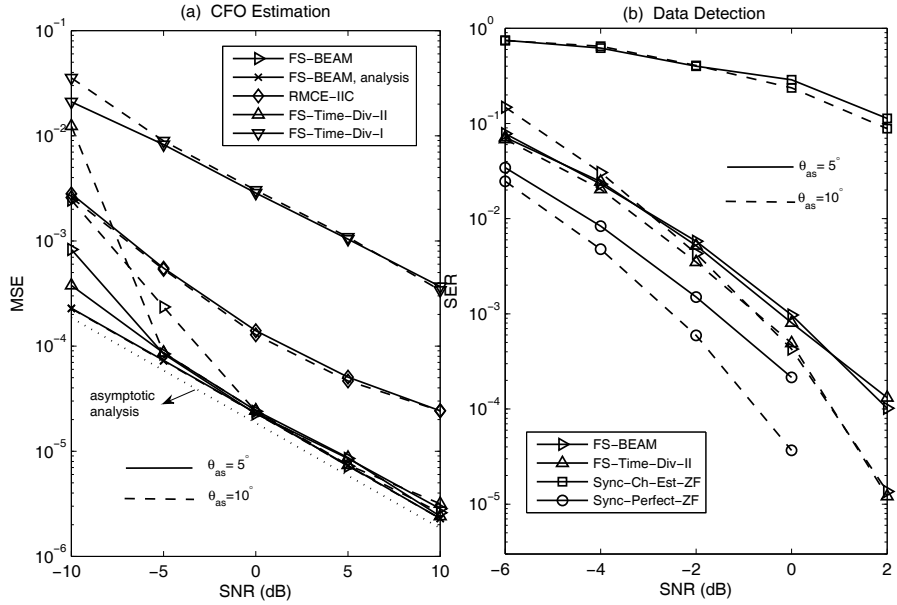}
\end{center}
\caption{
\textcolor{black}{
CFO estimation and SER performance of the proposed FS-BEAM scheme ($K=4$). }
}
\end{figure}

First, we consider the case of $K=4$ and plot the iterative procedure of CFO estimation in the proposed FS-BEAM under different SNR conditions in Fig. 3. The angular spread is $\theta_{\rm as}=5^\circ$ and the mean DOAs of the four users are fixed as $\{30^\circ, 60^\circ, 120^\circ, 150^\circ \}$. We consider both the cases of $\phi_{\max}=0.1$ and $\phi_{\max}=0.2$ in this example.
The CFOs are initialized as all zeros, i.e., $\hat\phi_k^{(0)}=0$, $k=1,2,\cdots,K$.
The results clearly demonstrate the fast convergence rate of the proposed CFO estimation method. Especially, in the case with relatively small maximum CFO, i.e., $\phi_{\max}=0.1$, two iterations are sufficient for convergence under the moderate SNR condition, while only a few more iterations are required for the low SNR condition or a larger maximum CFO. In the following, we consider $\phi_{\max}=0.2$ and five iterations are employed in FS-BEAM.

We show the CFO estimation performance of FS-BEAM with $K=4$ users in Fig. 4(a) being compared with RMCE-IIC.
The mean DOAs are set the same as  Fig. 3.
The corresponding analytical results from (\ref{MSEhatphik}) are plotted as the curves with marker `$\times$', \textcolor{black}{while the asymptotic analysis results computed from (\ref{MSEasym}) are represented by the dotted curve.}
Both two angular spreads $\theta_{\rm as}=5^\circ$ and $\theta_{\rm as}=10^\circ$ are included.
\textcolor{black}{We also include the following two time-division training based schemes for comparison, labelled as `FS-Time-Div-I' and `FS-Time-Div-II', as illustrated in Fig. 5.
\begin{itemize}
\item FS-Time-Div-I: Each user transmits a short training sequence of length-$N_{ s}$ ($N_{s}\!<\! N$) in a time-division manner at the beginning of uplink frame.  CP is also inserted in front of each sequence to eliminate the interference between the adjacent sequences in multipath propagation channel. After the $K$ time-division sequences, all users start data block transmission simultaneously. In FS-Time-Div-I, BS can perform single-user joint CFO and channel estimation for each user individually in the traditional way. The overall training length of FS-Time-Div-I can be expressed as $K(N_{\rm cp}+N_{ s})$.
\item FS-Time-Div-II:  At the beginning of uplink frame, each user first transmits time-division short training sequences of length-$N_s$ in the same way as FS-Time-Div-I.  Then, each user simultaneously transmits a length-$N$ training block as well as the data blocks. BS first perform DOA identification for each user individually by using the time-division training sequences. Denote ${\bf Y}_{k,{\rm TD}}$ as the received $N_{ s}\times M$ signal matrix corresponding to the short training sequence from the $k\rm{th}$ user. The DOA estimation for the $k\rm{th}$ user can be performed as $\hat\theta_{k,{\rm TD}} = \arg\max\limits_{\tilde\theta_k} \int_{\tilde\theta_k-\theta_{\rm as}}^{\tilde\theta_k+\theta_{\rm as}}\| {\bf Y}_{k,{\rm TD}} {\boldsymbol a}^*(\tilde\theta)\|^2  d\tilde\theta$.
    Then, similarly to (\ref{equ8}), to mitigate MUI for the $k\rm{th}$ user, BS can employ all steering vectors ${\boldsymbol a}(\tilde\theta)$ with $\tilde\theta\in (\hat\theta_{k,{\rm TD}} - \theta_{\rm as}, \hat\theta_{k,{\rm TD}} + \theta_{\rm as})$ as the receive beamforming vectors  on the subsequent length-$N$ training block as well as data blocks. Afterwards, the conventional single-user synchronization and channel estimation can be performed for each user based on the length-$N$ training block. The overall training length of FS-Time-Div-II can be expressed as $K(N_{\rm cp}+N_{ s})+ N_{\rm cp} + N$.
\end{itemize}
}

\begin{figure}[h]
\begin{center}
\includegraphics[width=130mm]{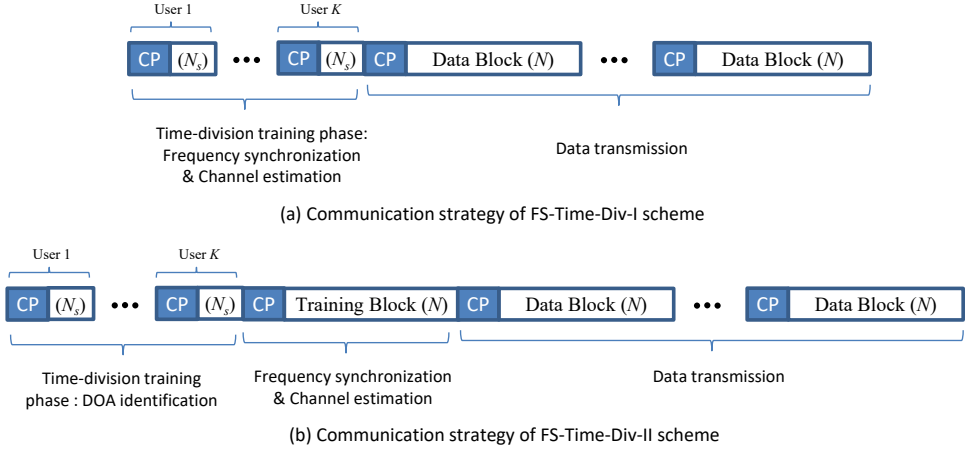}
\end{center}
\caption{\textcolor{black}{The uplink communication strategy with time-division training blocks. } }
\end{figure}

\begin{figure}[t]
\begin{center}
\includegraphics[width=120mm]{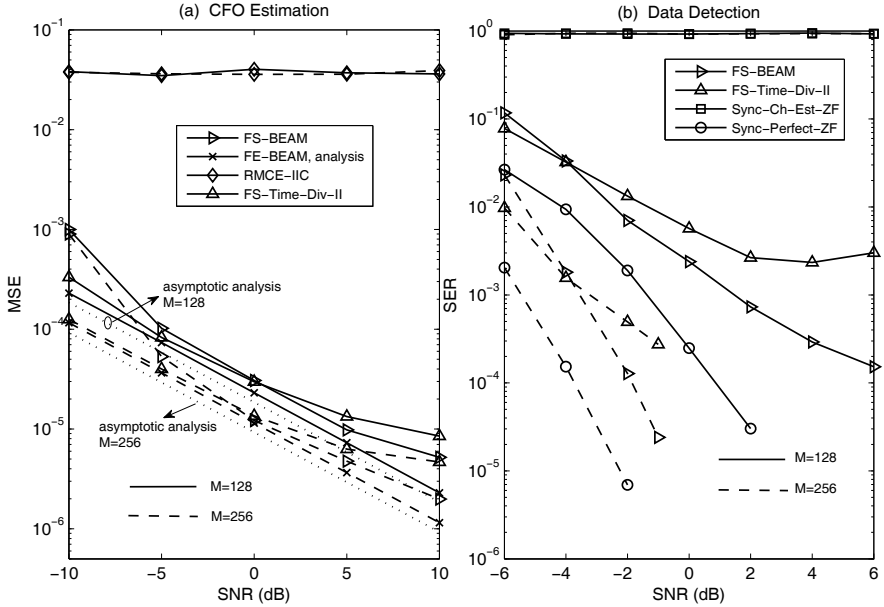}
\end{center}
\caption{
\textcolor{black}{
CFO estimation and SER performance of the proposed FS-BEAM scheme ($K=10$). }
}
\end{figure}

We consider $N_{ s}=16$ in this example. The following observations can be made from Fig. 4(a): 1) We see that the proposed method almost touches the corresponding analytical results above SNR $=0$ dB. This indicates that our estimation method can effectively suppress the MUI effect by exploiting the high spatial resolution from the large-scale antenna array. In this example, the performance of our method is basically unchanged with different magnitude of angular spreads, unless under the low SNR condition.
\textcolor{black}{ 2)  It is also seen that our method could behave much better
than RMCE-IIC and a performance gap as large as 10 dB is observed. Moreover, the proposed method also substantially outperforms FS-Time-Div-I scheme. This is not unexpected and because the training block in FS-BEAM could span a much wider time interval as compared to the time-division sequence for each user in FS-Time-Div-I, which provides the potential of higher resolution for CFO estimation~\cite{MinnTC05}. In the meanwhile, the proposed scheme could work similarly as FS-Time-Div-II.
 Since FS-Time-Div-II decouples the DOA identification and CFO estimation problems, it should have lower computational complexity than FS-BEAM.
Nevertheless, we should note that, the  training overhead of FS-Time-Div-II is about $1+\frac{K(N_{ s}+N_{\rm cp})}{N+N_{\rm cp}} = 237\%$ of the proposed scheme in this example. In other words, we may conclude that as compared to FS-Time-Div-II, the proposed scheme could remarkably  reduce the training overhead without obviously sacrificing the estimation performance.
}

The corresponding symbol error rate (SER) performance with $K=4$ users by using our proposed FS-BEAM scheme is depicted in Fig. 4(b). The result of FS-Time-Div-I is not included in this example due to its poor synchronization performance. \textcolor{black}{
 Moreover, for comparison, we also include the ideal SER results without the multiuser CFOs as the benchmark, referred to as `Sync-Perfect-ZF' and `Sync-Ch-Est-ZF'.
\textcolor{black}{Specifically, for the ideal benchmark schemes, we consider that all users are perfectly synchronized with BS in carrier frequency and conventional ZF data detection is employed.}
Moreover, BS has perfect knowledge of channel information in Sync-Perfect-ZF, while obtains channel information from conventional LS algorithm in Sync-Ch-Est-ZF. }
\textcolor{black}{
Note that after the initial frequency synchronization,
the residual frequency offset would lead to accumulative phase shift over the data blocks and may result in large detection error rate. Hence, phase tracking algorithms can be further employed to deal with this issue~\cite{Nikitopoulos05}.
For example, the authors in~\cite{Nikitopoulos05} have analytically evaluated  the effect of phase impairment due to phase noise plus a residual frequency offset and studied the corresponding compensation scheme.
However, in this work we mainly focus on the initial frequency synchronization performance, and thus, for simplicity we adopt only the first data block to evaluate the SER performance in simulations.}
The following observations can be made:
1) The results clearly demonstrate the validity of our proposed frequency synchronization scheme. \textcolor{black}{We also see that the SER curves of both Sync-Perfect-ZF and the proposed scheme drop more quickly with a larger angular spread under high SNR condition.
This phenomenon can be explained as follows. A larger angular spread may increase the rank of the channel covariance matrix from each user~\cite{Adhikary13}, and thus can potentially provide a higher diversity gain of data detection performance~\cite{Tse}.
}  2) It is seen that the performance gap between our scheme and Sync-Perfect-ZF is less than 2 dB with $K=4$ users.
   Our scheme can substantially outperform Sync-Ch-Est-ZF. This is not unexpected and
because Sync-Ch-Est-ZF does not first perform user separation from spatial domain and
\textcolor{black}{ needs to estimate a total of $KL=40$ channel parameters based on the $N=64$ received training samples.
The insufficiency of training samples of Sync-Ch-Est-ZF would lead to both poor channel estimation and detection performance.
3) It is observed that SER curve of the proposed scheme basically touches the corresponding results of FS-Time-Div-II. This implies that as compared to FS-Time-Div-II, the proposed scheme could remarkably reduce the training overhead without obvious sacrifice on detection performance.}

\textcolor{black}{
Next, we evaluate the CFO estimation and SER performance in Fig. 6 with $K=10$ users and $\theta_{\rm as}=5^\circ$.
The mean DOAs of ten users are fixed as $\{
30^\circ, 45^\circ, 120^\circ, 150^\circ, 80^\circ, 95^\circ,  20^\circ, 60^\circ, 110^\circ, 130^\circ\}$. We include both results of $M=128$ and $M=256$ in this example.
The user separation  in spatial domain becomes smaller with $K=10$ and
thus lead to the increased MUI effect.
We can observe some discrepancy between FS-BEAM and the corresponding analytical results under high SNR from Fig. 6(a).
As expected, more receive antennas at BS bring higher spatial resolution and thus improve the system performance in both CFO estimation and data detection.
Especially, FS-BEAM shows some observable SER performance floor with $M=128$, while it would work far more better when $M$ increases to $256$. Moreover, we see that FS-BEAM  outperform FS-Time-Div-II under moderate and high SNR condition. This should be attributed to the adaptive beamforming scheme in FS-BEAM, which employs only the qualified DOAs for beamforming. In comparison,
FS-Time-Div-II employs all DOAs within the angular spread of each user for beamforming, which suffers from a certain magnitude of MUI effect with $K=10$ users.
}

\begin{figure}[t]
\begin{center}
\includegraphics[width=120mm]{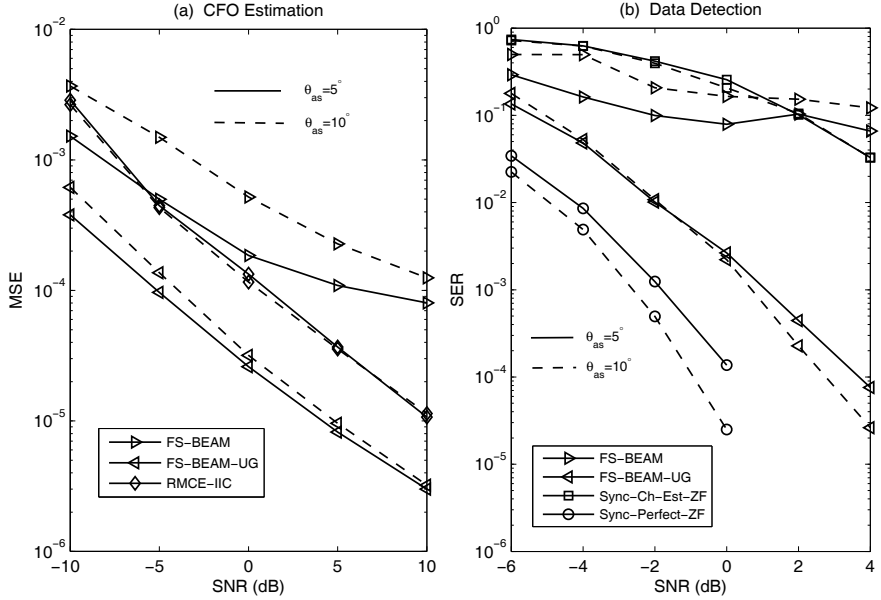}
\end{center}
\caption{
\textcolor{black}{
CFO estimation and SER performance of the proposed schemes with randomly distributed DOAs ($K=4$).}
}
\end{figure}

\begin{figure}[h]
\begin{center}
\includegraphics[width=120mm]{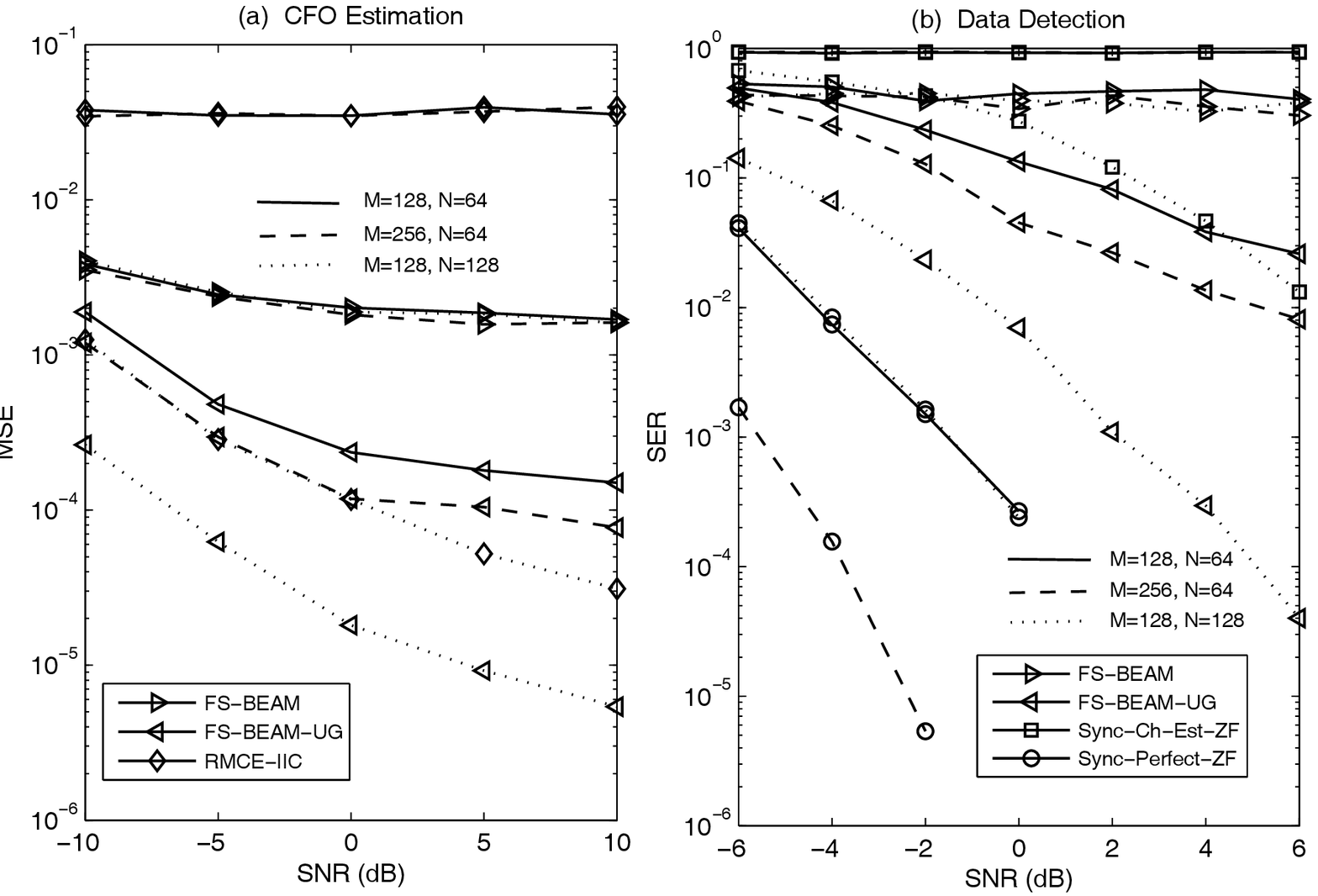}
\end{center}
\caption{
\textcolor{black}{
CFO estimation and SER performance of the proposed schemes with randomly distributed DOAs ($K=10$).}
}
\end{figure}

\begin{figure}[t]
\begin{center}
\includegraphics[width=100mm]{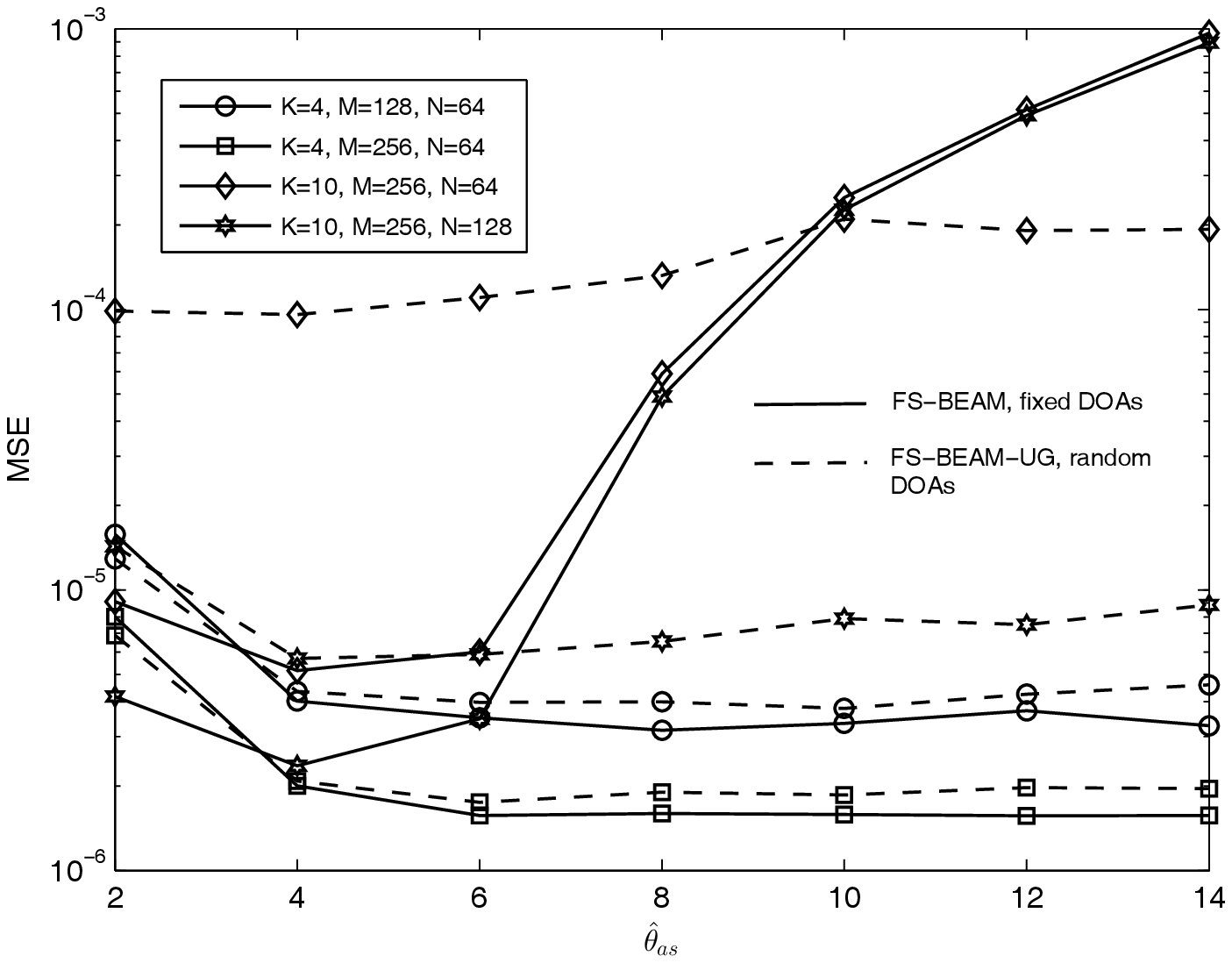}
\end{center}
\caption{
\textcolor{black}{
CFO estimation performance of the proposed schemes with imperfect knowledge of angular spread. }
}
\end{figure}

Considering the user locations are randomly distributed, we next assume the mean DOA of each user is randomly generated in each simulation trial. Specifically, we assume the mean DOAs of half of the users follow uniform distribution from $30^\circ$ to $60^\circ$ where the rest half follow uniform distribution from $120^\circ$ to $150^\circ$.
  We plot the CFO estimation performance with $K=4$ of FS-BEAM and FS-BEAM-UG in Fig. 7(a). Both angular spreads of $\theta_{\rm as}=5^\circ$ and $\theta_{\rm as}=10^\circ$ are considered in this example.
 From the results, we can observe a performance error floor of the proposed FS-BEAM, especially with larger angular spread. This is not unexpected and mainly due to the cases where some users have very similar mean DOAs and the spatial separation may not be enough.
Nevertheless, it is observed that our proposed FS-BEAM-UG could work with randomly distributed users and can achieve much better performance as compared to FS-BEAM, which indicates the robustness of FS-BEAM-UG. Moreover, we see that the proposed FS-BEAM-UG also substantially outperforms the competitor of RMCE-IIC and a performance gap more than 5 dB can be observed.
The corresponding SER performance with $K=4$ is depicted in Fig. 7(b).
 Both the ideal benchmarks of Sync-Perfect-ZF and Sync-Ch-Est-ZF are included for comparison. As expected, the proposed FS-BEAM is sensitive to the random distribution of user DOAs and cannot work properly in this example.
This suggests that our FS-BEAM may rely on some proper user schedule strategy to avoid the case where some users do not have enough spatial separation.
 Nevertheless, we see that our proposed FS-BEAM-UG still would work properly and  substantially outperform both the Sync-Ch-Est-ZF and FS-BEAM scheme.

\textcolor{black}{
Next, we evaluate the CFO estimation and SER performance of $K=10$ users  with randomly distributed DOAs in Fig. 8. We include additional results of $M=256$ and $N=128$ in this example. The angular spread is assumed as $\theta_{\rm as}=5^\circ$. As expected,
FS-BEAM shows little change when $M$ or $N$ increases due to the severe MUI effect.
In comparison, we can observe performance improvement of the proposed FS-BEAM-UG by increasing $M$ or $N$. Specifically, we can still observe error performance floor of FS-BEAM-UG in terms of both CFO estimation and data detection with $N=64$ even when $M$ increases to 256, whereas it can behave much better when $N$ increases to 128. This is because FS-BEAM-UG should perform joint estimation for the grouped users and the training sequence length becomes a bottleneck in this situation.
Moreover,  Fig. 8(a) shows that, with a larger training block length of $N=128$, the CFO estimation MSE of RMCE-IIC could also decrease as SNR increases. Nevertheless, the performance gap between RMCE-IIC and the proposed FS-BEAM-UG is still more than 10 dB.
}

In the next example, we investigate the sensitivity of the proposed scheme with respect to the imperfect knowledge of angular spread. Suppose BS takes the angular spread as $\hat\theta_{\rm as}$. In Fig. 9, we evaluate the impact of the assumed angular spread $\hat\theta_{\rm as}$ on the CFO estimation performance under SNR $=5$ dB.
We consider $\theta_{\rm as}=5^\circ$ in this example. \textcolor{black}{ Both the results of fixed DOAs and random DOAs are considered, represented by the solid and dashed curves, respectively. Specifically, the fixed DOAs are  generated in the same way as Fig. 4 and Fig. 6 for $K=4$ and $K=10$, respectively.  The random DOAs are generated in the same way as Fig. 7 and Fig. 8.
 We can observe that with $K=4$ users,
  only a little performance degradation when the assumed angular spread $\hat\theta_{\rm as}$ deviates from the true value, which says that our schemes are robust to the imperfect knowledge of angular spread in this situation.
 However, from the results with $K=10$ users, we see that the proposed FS-BEAM scheme shows much more sensitivity to the imperfect knowledge of angular spread. This is not unexpected and  because more users would lead to fewer `white space' in spatial domain and decrease the robustness to imperfect knowledge of angular spread. }

\begin{figure}[t]
\begin{center}
\includegraphics[width=100mm]{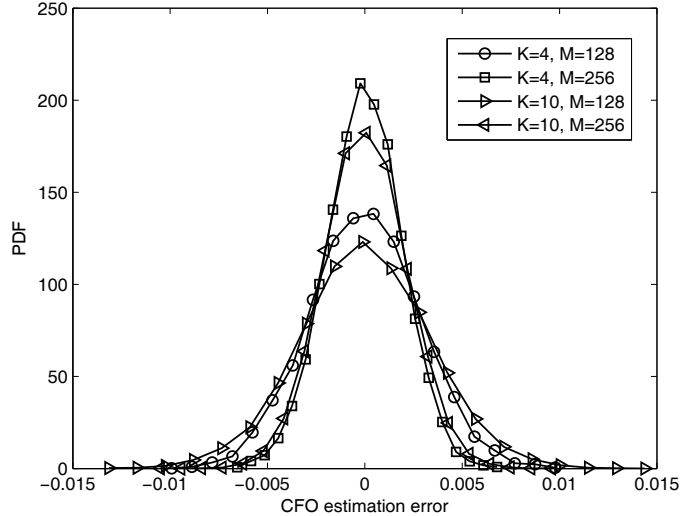}
\end{center}
\caption{
\textcolor{black}{PDF of CFO estimation of the proposed FS-BEAM scheme.
} }
\end{figure}

\begin{figure}[t]
\begin{center}
\includegraphics[width=100mm]{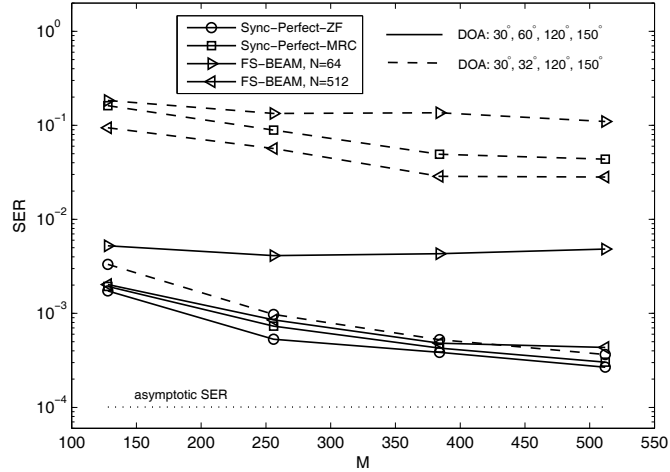}
\end{center}
\caption{ \textcolor{black}{ SER performance of the proposed schemes as a function of $M$. }
}
\end{figure}

\textcolor{black}{In Fig. 10, we plot the probability density function (PDF) of the CFO estimation error of the proposed FS-BEAM with different number of receive antennas. In this example, we include both the cases for $K=4$ and $K=10$, where the DOAs are generated in the same way as Fig. 4 and Fig. 6, respectively. As expected, the error distribution tends to be more concentrated to zero with more receive antennas. }

\textcolor{black}{It is also interesting to evaluate the performance of the proposed method as $M$ increases. In Fig. 11, we plot the SER performance of the proposed FS-BEAM scheme with $K=4$ users. We consider the following two different DOA configurations: $\{30^\circ, 60^\circ, 120^\circ, 150^\circ \}$ and $\{30^\circ, 32^\circ, 120^\circ, 150^\circ \}$, represented by the solid and dashed curves, respectively.  The overall SNR of the multiple receive antennas can be expressed as $M\sigma_s^2/\sigma_n^2$, which is the ratio between the overall average received signal power at multiple receive antennas and the noise power.
The aforementioned  SNR defined as $\sigma_s^2/\sigma_n^2$ can be interpreted as the SNR per receive antenna, since the average channel gain from each user at one receive antenna is normalized, i.e., $\sum_{l=1}^L\sigma_{h,l}^2=1$.
 In this example, we vary the number of antennas with fixed overall SNR $M\sigma_s^2/\sigma_n^2=19$ dB.
 The performance of maximum ratio combining (MRC) detection with perfect knowledge of channel information and perfect synchronization is also plotted for comparison, labelled as `Sync-Perfect-MRC'.
The following observations can be made:
1) Regarding the solid curves, since each user has enough spatial separation from the others, we see that Sync-Perfect-MRC closely approaches the corresponding performance of Sync-Perfect-ZF, especially as $M$ increases. Moreover, note that SNR $\sigma_s^2/\sigma_n^2$ in fact decreases when $M$ increases with fixed overall SNR $M\sigma_s^2/\sigma_n^2$. The proposed FS-BEAM exhibits no much SER performance improvement as $M$ increases with a smaller training sequence length $N$.
Nevertheless, with enough large $N$, the proposed FS-BEAM also approaches Sync-Perfect-ZF.
2) In comparison, from the dashed curves, we see that due to the insufficient spatial separation, the performance of both Sync-Perfect-MRC and FS-BEAM suffers from substantial deterioration and becomes far away from that of Sync-Perfect-ZF.
3) We also plot the single-user analytical SER~\cite[eq. (45)]{SimonIEEE} under SNR= 19 dB and AWGN channel as a benchmark, which can be considered as the asymptotic performance when $M$ tends to be infinity. Owing to the MUI interference as well as the channel fading effect,  a performance gap between Sync-Perfect-ZF and the asymptotic benchmark is still observable.
}

Let us now evaluate the computational complexity of the proposed CFO estimation in terms of complex multiplications. We denote $\alpha$ as the number of required iterations in FS-BEAM. Then, the complexity of the proposed FS-BEAM is approximately in the order of $\mathcal{O}\big(  NM_{\rm FFT} \log_2 M_{\rm FFT} + 2\alpha KL NM_{\rm FFT} \big) \simeq  \mathcal{O}\big(  2\alpha KL NM_{\rm FFT} \big)$. In addition,
\textcolor{black}{considering the multiuser estimation in one group with user number of $\kappa$ in FS-BEAM-UG,}
SVD on $\mathbb{B}(\hat{\boldsymbol\phi})$ requires around $\mathcal{O}(\kappa^2 L^2N )$, and computing ${\bf R}_{\rm Grp}$ requires $\mathcal{O}(\kappa L N^2 )$.
Calculating (\ref{equ30})
requires around $\mathcal{O}\big(  \kappa L N^2+  2 \kappa^2 LN^2 + 2\kappa N^3 \big)$. Then, the total complexity for this group can be expressed as
$\mathcal{O}\big( 2\kappa N^3 + 2(\kappa^2+\kappa)LN^2 + \kappa^2L^2 N \big)\simeq \mathcal{O}(4\kappa N^3)$. Considering the extreme case where all users are grouped, as compared to FS-BEAM, the FS-BEAM-UG further requires the complexity of $\mathcal{O}(4KN^3)$.
On the other side, the complexity of RMCE-IIC can be expressed as $\mathcal{O}(\eta \beta  K LNM)$, where $\eta$ is the required iterations and $\beta$ denotes the number of trial CFOs in~\cite[eq. (21)]{Tsai13}. Let us consider the example of above $K=6$ scenario. The required complexities of the proposed FS-BEAM, FS-BEAM-UG and RMCE-IIC can be approximately expressed as $\mathcal{O}(5.9\times 10^6)$, $\mathcal{O}(1.2\times 10^7)$ and $\mathcal{O}(1.8\times 10^8)$, respectively. This indicates that the proposed schemes also have the advantage of lower complexity as compared to RMCE-IIC.

\section{Conclusions}

In this paper, we developed a new frequency synchronization scheme for massive multiuser uplink transmissions by exploiting the angle domain information of different users. We designed both the CFO estimation and beamforming method, and convert the original complicated multi-user problem into equivalent single user transmission model. We further design an improved user grouping scheme to deal with the unexpected scenarios that some users may not be separated well from the spatial domain. Both the simulation and theoretical results are provided to corroborate the proposed studies.

\appendices

\section{Asymptotic CFO Estimation Performance}

\textcolor{black}{We have
${\bf R}_k^2  =   \frac{1}{4\theta_{\rm as }^2}   \int_{\theta_k-\theta_{\rm as}}^{\theta_k+\theta_{\rm as}}   \int_{\theta_k-\theta_{\rm as}}^{\theta_k+\theta_{\rm as}}  {\boldsymbol a}^*(\tilde\theta_1)
   {\boldsymbol a}^T(\tilde\theta_1)  {\boldsymbol a}^*(\tilde\theta_2)
   {\boldsymbol a}^T(\tilde\theta_2)
   d\tilde\theta_1  d\tilde\theta_2$.
With small angular spread $\theta_{\rm as}$, we further obtain
\begin{align}
& \int_{\theta_k-\theta_{\rm as}}^{\theta_k+\theta_{\rm as}}
{\boldsymbol a}^T(\tilde\theta_1)  {\boldsymbol a}^*(\tilde\theta_2)  {\boldsymbol a}^T(\tilde\theta_2)
d\theta_2 \simeq   \int_{\theta_k-\theta_{\rm as}}^{\theta_k+\theta_{\rm as}}
   \frac{ \sin \frac{\chi M(\cos\theta_1-\cos\theta_2)}{2}  }{ \frac{\chi (\cos\theta_1-\cos\theta_2)}{2}} {\rm e}^{-{\bf j}\frac{\chi(M-1)(\cos\theta_1- \cos\theta_2)}{2}}
  {\boldsymbol a}^T(\tilde\theta_2)
d\tilde\theta_2  \nonumber \\
\simeq &  \frac{1}{\sin\theta_k}\int_{\cos(\theta_k+\theta_{\rm as}) }^{\cos(\theta_k-\theta_{\rm as}) }
   \frac{ \sin \frac{\chi M(\cos\tilde\theta_2-\cos\tilde\theta_1)}{2}  }{ \frac{\chi (\cos\tilde\theta_2-\cos\tilde\theta_1)}{2}} {\rm e}^{-{\bf j}\frac{\chi(M-1)(\cos\tilde\theta_2- \cos\tilde\theta_1)}{2}}
  {\boldsymbol a}^T(\tilde\theta_2)
d \cos\tilde\theta_2 \nonumber \\
\simeq & \frac{2\pi }{\chi \sin\theta_k}  {\boldsymbol a}^T(\tilde\theta_1). \label{equappendix}
\end{align}
The derivation of (\ref{equappendix}) arises from the following approximation: For any function $f(\cdot)$ and $a>0$, $b>0$, there holds
$\lim\limits_{M\to \infty} \int_{x_0-a}^{x_0+b} \frac{\sin \frac{\chi M (x-x_0)}{2}}{\chi (x-x_0)/2} f(x) dx \simeq \lim\limits_{M\to \infty} \int_{-\infty}^{\infty} \frac{\sin \frac{\chi M (x-x_0)}{2}}{\chi (x-x_0)/2} f(x) dx  = \frac{2\pi}{\chi}f(x_0)$. Then, based on (\ref{equappendix}), we can rewrite ${\bf R}_k^2$ as
\begin{align}
{\bf R}_k^2 \simeq \frac{1}{4\theta_{\rm as}^2} \frac{2\pi }{\chi \sin\theta_k}
\int_{\theta_k-\theta_{\rm as}}^{\theta_k+\theta_{\rm as}}
{\boldsymbol a}^*(\tilde\theta_1)
   {\boldsymbol a}^T(\tilde\theta_1)
d\tilde\theta_1 = \frac{\pi}{ \theta_{\rm as}\chi\sin\theta_k } {\bf R}_k.
\end{align}
Hence, with  sufficiently large $M$, we have the approximation:
$\textrm{Tr}( {\bf R}_k^2 ) \simeq \frac{\pi }{\theta_{\rm as} \chi\sin\theta_k } \textrm{Tr}( {\bf R}_k ) \frac{M \pi }{\theta_{\rm as} \chi\sin\theta_k }$ and $\textrm{Tr}( {\bf R}_k^3 )
 = \frac{\pi^2 }{\theta_{\rm as}^2  \chi^2 \sin\theta_k^2 } \textrm{Tr}( {\bf R}_k ) = \frac{M \pi^2 }{\theta_{\rm as}^2  \chi^2 \sin\theta_k^2 }$.
 Then, the asymptotic CFO estimation MSE with sufficiently large $M$ for the $k\rm{th}$ user can be expressed as
  \begin{align}\label{appequ1}
  \textrm{MSE}_{asym,M}\{\hat\phi_k\} =  \frac{  \sigma_n^2   }{ 2 M\cdot \| {\bf P}_{{\bf B}_k}^\bot{\bf D}^H {\bf B}_k {\bf\Sigma}_h\|^2  }.
  \end{align} }
\textcolor{black}{
Next, with sufficiently large $N$, we have the approximation: ${\bf B}^H{\bf B} \simeq N\sigma_s^2 {\bf I}_L $. Then, there holds
\begin{align}\label{appequ2}
\| {\bf P}_{{\bf B}_k}^\bot{\bf D}^H {\bf B}_k {\bf\Sigma}_h\|^2 \simeq &
\textrm{Tr}({\bf\Sigma}_h^H {\bf B}_k^H {\bf D} {\bf D}^H {\bf B}_k {\bf\Sigma}_h  )
 -
 \frac{1}{N\sigma_s^2}
 \textrm{Tr}(
 {\bf\Sigma}_h^H {\bf B}_k^H {\bf D} {\bf B}_k{\bf B}_k^H  {\bf D}^H {\bf B}_k {\bf\Sigma}_h  ) \nonumber \\
 \simeq & \sigma_s^2 \textrm{Tr}(  {\bf D}{\bf D}^H )  \textrm{Tr}( {\bf\Sigma}_h^H{\bf\Sigma}_h ) - \frac{\sigma_s^2}{N}  \textrm{Tr}({\bf D}) \textrm{Tr}({\bf D}^H)   \textrm{Tr}( {\bf\Sigma}_h^H{\bf\Sigma}_h ) \nonumber\\
 = &  \sigma_s^2 \textrm{Tr}(  {\bf D}{\bf D}^H )  - \frac{\sigma_s^2}{N}  \textrm{Tr}({\bf D}) \textrm{Tr}({\bf D}^H)   \simeq \frac{\pi^2 N \sigma_s^2}{3}.
\end{align}
After substituting (\ref{appequ2}) into (\ref{appequ1}),  we arrive at (\ref{MSEasym}).
}

\linespread{1.0}


\begin{thebibliography}{99}


\bibitem{Marzetta10}
T. Marzetta, ``Noncooperative cellular wireless with unlimited
numbeers of base station antennas,'' \emph{IEEE Trans. Wireless
Commun.}, vol. 9, no. 11, pp. 3590--3600, Nov. 2010.

\bibitem{Choi14}
J. Choi, D. J. Love, and P. Bidigare, ``Downlink training techniques for FDD massive MIMO systems: open-loop and closed loop training with memory,'' \emph{IEEE J. Sel. Topics Signal Process.}, vol. 8, no. 5, pp. 802--814, Oct. 2014.


\bibitem{RusekSPM}
F. Rusek, D. Persson, B. K. Lau, E. G. Larsson, T. L. Marzetta, O. Edfors, and F. Tufvesson, ``Scaling up MIMO: opportunities and challenges with very large arrays,'' \emph{IEEE Signal Process. Mag.,} vol. 30, no. 1, pp. 40--60, Jan. 2013.


\bibitem{LarssonCM}
E. Larsson, O. Edfors, F. Tufvesson, and T. Marzetta, ``Massive MIMO for next generation wireless systems,'' \emph{IEEE Commun. Mag.,} vol. 52, no. 2, pp. 186--195, Feb. 2014.




\bibitem{Adhikary13}
A. Adhikary, J. Nam, J.-Y. Ahn, and G. Caire, ``Joint spatial division and multiplexing--the large-scale array regime,'' \emph{IEEE Trans. Inf. Theory,} vol. 59, no. 10, pp. 6441--6463, Oct. 2013.

\bibitem{You15}
L. You, X. Gao, X. Xia, N. Ma, and Y. Peng, ``Pilot reuse for massive MIMO transmission over spatially correleated rayleigh fading channels,'' vol. 14, no. 6, pp. 3352--3366, June 2015.

\bibitem{Sun15}
C. Sun, X. Gao, S. Jin, M. Matthaiou, Z. Ding, and C. Xiao, ``Beam division multiple access transmission for massive MIMO communications,'' \emph{IEEE Trans. Commun.}, vol. 63, no. 6, pp. 2170--2184, June 2015.


\bibitem{Schmidl97}
T. M. Schmidl and D. C. Cox, ``Robust frequency and timing synchronization for OFDM,'' \emph{IEEE Trans. Commun.,} vol. 45, no. 12, pp. 1613--1621, Dec. 1997.

\bibitem{Minn03}
H. Minn, V. K. Bhargava, and K. B. Letaief, ``A robust timing and frequency synchronization for OFDM systems,'' \emph{IEEE Trans. Wirel. Commun.}, vol. 2, no. 4, pp. 822--839, Jul. 2003.


\bibitem{Yao05}
Y. Yao and G. B. Giannakis, ``Blind carrier frequency offset
estimation in SISO, MIMO, and multiuser OFDM systems,'' \emph{IEEE
Trans. Commun.}, vol. 53, no. 1, pp. 173--183, Jan. 2005.



\bibitem{Lmai:TVT14}
S. Lmai, A. Bourre, C. Laot, and S. Houcke, ``An efficient blind
estimation of carrier frequency offset in OFDM systems,'' \emph{IEEE
Trans. Veh. Technol.}, vol. 63, no. 4, pp. 1945--1950, May 2014.

\bibitem{ZhangTSP14}
W. Zhang, Q. Yin, W. Wang, and F. Gao, ``One-shot blind CFO and
channel estimation for OFDM with multi-antenna receiver,''
\emph{IEEE Trans. Signal Process.}, vol. 62, no. 15, pp. 3799--3808,
Aug. 2014.


\bibitem{ZhangTWC16}
W. Zhang, Q. Yin, and F. Gao, ``Computationally efficient blind estimation of carrier frequency Offset for MIMO-OFDM Systems,'' \emph{IEEE Trans. Wireless Commun.}, accepted for publication, 2016.



\bibitem{Besson}
O. Besson and P. Stoica, ``On parameter estimation of MIMO
flat-fading channels with frequency offsets,'' \emph{IEEE Trans.
Signal Process.}, vol. 51, no. 3, pp. 602--613, 2003.



\bibitem{JChen08}
J. Chen, Y. C. Su, S. Ma, and T. S. Ng, ``Joint CFO and channel
estimation for multiuser MIMO-OFDM systems with optimal training
sequences,'' \emph{IEEE Trans. Signal Process.}, vol. 56, no. 8, pp.
4008--4019, 2008.

\bibitem{YWuEURAPSIP}
Y. Wu, J. W. M. Bergmans, and S. Attallah, ``Carrier frequency offset
estimation for multiuser MIMO OFDM uplink using CAZAC sequences:
performance and sequence optimization,'' \emph{EURASIP Journal on
Wirel. Commun. and Network.}, vol. 2011, Article ID: 570680.


\bibitem{Tsai13}
Y.-R. Tsai, H.-Y. Huang, Y.-C. Chen, and K.-J. Yang, ``Simultaneous multiple carrier frequency offsets estimation for coordinated multi-point transmission in OFDM systems,'' \emph{IEEE Trans. Wirel. Commun.}, vol. 12, no. 9, Sept. 2013.


\bibitem{DuttaTVT15}
A. K. Dutta, K. V. S. Hari, and L. Hanzo, ``Minimum-error-probability CFO estimation for multiuser MIMO-OFDM systems,'' \emph{IEEE Trans. Veh. Technol.,} vol. 64, no. 7, pp. 2804--2818, Jul. 2015.




\bibitem{Cheng13}
H. V. Cheng and E. G. Larsson, ``Some fundamental limits on frequency synchronization in massive MIMO,'' in Proc. \emph{IEEE Asilomar}, 2013.

\bibitem{MukherjeeTVT}
S. Mukherjee, S. K. Mohammed, and I. Bhushan, ``Impact of CFO estimation on the performance of ZF reciever in massive MU-MIMO systems,'' \emph{IEEE Trans. Veh. Technol.,}  2016, DOI: 10.1109/TVT.2016.2518401.


\bibitem{Mukherjee15}
S. Mukherjee and S. K. Mohammed, ``Low-complexity CFO estimation for multi-user massive MIMO systems,'' in \emph{Proc. IEEE GLOBECOM}, 2015.
\textcolor{black}{
\bibitem{MukherjeeGLO16}
S. Mukherjee and S. K. Mohammed, ``Constant envelope pilot-based low-complexity CFO estimation in massive MU-MIMO systems,'' in \emph{Proc. IEEE GLOBECOM}, 2016.
\bibitem{Mukherjee16}
S. Mukherjee and S. K. Mohammed, ``Impact of frequency selectivity on the information rate performance of CFO impaired single-carrier massive MU-MIMO uplink,'' \emph{IEEE Wirel. Commun. Lett.}, vol. 5, no. 6, pp. 648--651, Dec. 2016.}


\bibitem{Zhang15sub}
W. Zhang and F. Gao, ``Blind frequency synchronization for multiuser OFDM uplink with large number of receive antennas,''\emph{IEEE Trans. Signal Processing}, vol. 64, no. 9, 2016.



\bibitem{Gaoarxiv}
H. Xie, F. Gao, S. Zhang, S. Jin, ``A unified transmission strategy for TDD/FDD massive MIMO systems with spatial basis expansion model,'' to appear in
\emph{IEEE Trans. Veh. Technol.}, 2016.


\bibitem{Morelli}
M. Morelli, C.-C. J. Kuo, and M.-O. Pun, ``Synchronization techniques for orthogona frequency division multiple access (OFDMA): a tuturial review,'' in \emph{Proc. IEEE}, vol. 95, no. 7, Jul. 2007.

\bibitem{Sun09}
P. Sun and L. Zhang, ``Low complexity pilot aided frequency synchronization for OFDMA uplink transmissions,'' \emph{IEEE
Trans. Wireless Commun.}, vol. 8, no. 7, pp. 3758¨C3769, Jul. 2009.



\bibitem{Wilkinson}
J. H. Wilkinson, \emph{The Algebraic Eigenvalue Problem}, Oxford, U.K., Clarendon, 1965.


\bibitem{ZhangGLO16}
W. Zhang, F. Gao, and H. Wang, ``Frequency synchronization for massive MIMO multi-user uplink,'' in Proc. \emph{IEEE GLOBECOM}, 2016, to appear.
\textcolor{black}{
\bibitem{Nikitopoulos05}
K. Nikitopoulos and A. Polydoros, ``Phase-impairment effects and compensation algorithms for OFDM systems,'' \emph{IEEE Trans. on Commun.}, vol. 53, no. 4, pp. 698-707, Apr. 2005.
\bibitem{SimonIEEE}
M. K. Simon and M.-S. Alouini, ``A unified approach to the performance analysis of digital communication over generalized fading channels,'' in \emph{Proc. IEEE}, vol. 86, no. 9, pp. 1860-- 1877, Sept. 1998.
\bibitem{Tse}
D. Tse and P. Viswanath, ``Fundamentals of wireless communications,'' \emph{Cambridge University Press}, 2004.
\bibitem{Yin13}
H. Yin, D. Gesbert, M. filippou, and Y. Liu, ``A coordinated approach to channel estimation in large-scale multiple-antenna systems,'' \emph{IEEE Jounal on Selected Areas in Commun.}, vol. 31, no. 2, pp. 264--273, Feb. 2013.
\bibitem{GaoAccess}
H. Xie, F. Gao, and S. Jin, ``An overview of low-rank channel estimation for massive MIMO systems,'' \emph{IEEE Access}, vol. 4, pp. 7313--7321, Nov. 2016.
\bibitem{MinnTC05}
H. Minn and S. Xing, ``An optimal training signal structure for frequency offset estimation,'' \emph{IEEE Trans. on Commun.}, vol. 53, no. 2, pp. 343-355, Feb. 2005.
}


\end{thebibliography}
\end{document}